\documentclass[10pt,journal]{IEEEtran}

\usepackage{array, calc, color, multicol}
\usepackage{algorithm}
\usepackage[noend]{algorithmic} 
\usepackage{graphics, epstopdf, graphicx, epsfig, psfrag, epsf, subfigure}
\usepackage{amssymb, amsfonts, amsmath, times}
\usepackage{relsize}
\usepackage{multicol,balance, verbatim}
\usepackage{textcomp, mathrsfs, stmaryrd, setspace}
\usepackage{amssymb}
\usepackage{cite}

\newcommand{\ie}{{\em i.e., }}
\newcommand{\Ie}{{\em I.e., }}
\newcommand{\eg}{{\em e.g., }}

\newcommand{\ncrn}{{\tt{NCMI-Batch}}}
\newcommand{\ncin}{{\tt{NCMI-Instant}}}

\newtheorem{theorem}{Theorem}
\newtheorem{lemma}[theorem]{Lemma}

\newtheorem{corollary}[theorem]{Corollary}
\newtheorem{definition}{Definition}
\newtheorem{example}{Example}

\DeclareGraphicsExtensions{.eps, .pdf, .png, .jpg}   

\newcommand{\Nset}{\mathcal{N}}
\newcommand{\Mset}{\mathcal{M}}
\newcommand{\Hset}{\mathcal{H}}

\newcommand{\Wset}{\mathcal{W}}

\IEEEoverridecommandlockouts

\begin{document}
\title{Device-to-Device Networking Meets Cellular via Network Coding}
\author{Yasaman~Keshtkarjahromi,~\IEEEmembership{Student~Member,~IEEE,}
        Hulya~Seferoglu,~\IEEEmembership{Member,~IEEE,}
        Rashid~Ansari,~\IEEEmembership{Fellow,~IEEE,}
        and~Ashfaq~Khokhar,~\IEEEmembership{Fellow,~IEEE}
\IEEEcompsocitemizethanks{\IEEEcompsocthanksitem Y. Keshtkarjahromi, H. Seferoglu, and R. Ansari are with the Department of Electrical and Computer Engineering, University of Illinois at Chicago, Chicago, IL, 60607.
E-mail: ykesht2@uic.edu, hulya@uic.edu, ransari@uic.edu. A. Khokhar is with the Department of Electrical and Computer Engineering, Iowa State University, Ames, Iowa 50011. E-mail: ashfaq@iastate.edu.
}
\thanks{This work was partially supported by NSF grant CNS-0910988. The preliminary results of this paper were presented in part at the IEEE International Conference for Military Communications (MILCOM), Tampa, FL, Oct. 2015.}
}

\maketitle

{$\hphantom{a}$}\vspace{-10pt}{}
	
\begin{abstract}
Utilizing device-to-device (D2D) connections among mobile devices is promising to meet the increasing throughput demand over cellular links. In particular, when mobile devices are in close proximity of each other and are interested in the same content, D2D connections such as Wi-Fi Direct can be opportunistically used to construct a cooperative (and jointly operating) cellular and D2D networking system. However, it is crucial to understand, quantify, and exploit the potential of network coding for cooperating mobile devices in the joint cellular and D2D setup. In this paper, we consider this problem, and (i) develop a network coding framework, namely NCMI, for cooperative mobile devices in the joint cellular and D2D setup, where cellular and D2D link capacities are the same, and (ii) characterize the performance of the proposed network coding framework, where we use packet completion time, which is the number of transmission slots to recover all packets, as a performance metric. We demonstrate the benefits of our network coding framework through simulations.
\end{abstract}

\vspace{-10pt}
\section{\label{sec:introduction} Introduction}
\vspace{-5pt}
The increasing popularity of diverse applications in today's mobile devices introduces higher demand for throughput, and puts a strain especially on cellular links. In fact, cellular traffic is growing exponentially, and it is expected to remain so for the foreseeable future \cite{cisco_index, ericsson_report}.

\begin{figure}[t!]
\centering
\subfigure[The default operation to connect to the Internet]{ \scalebox{.18}{\includegraphics{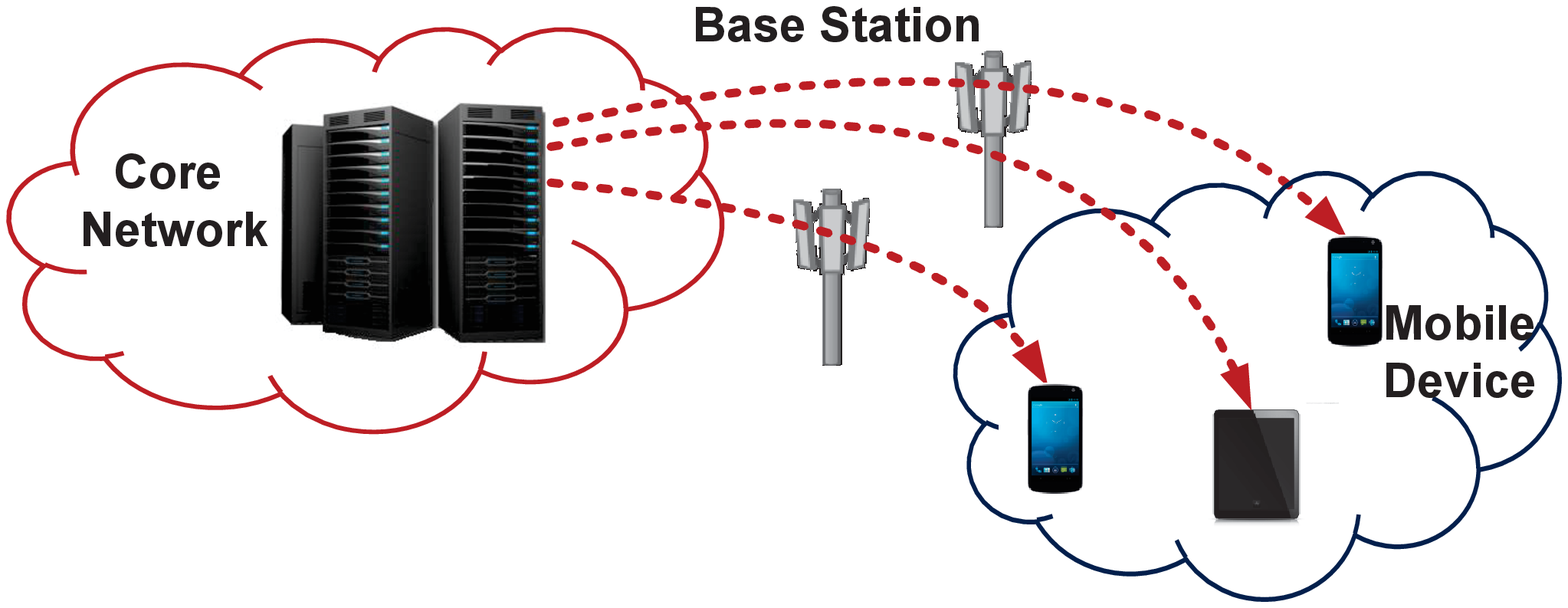}} } \hspace{5pt}
\subfigure[Using both cellular and D2D interfaces simultaneously ]{ \scalebox{.18}{\includegraphics{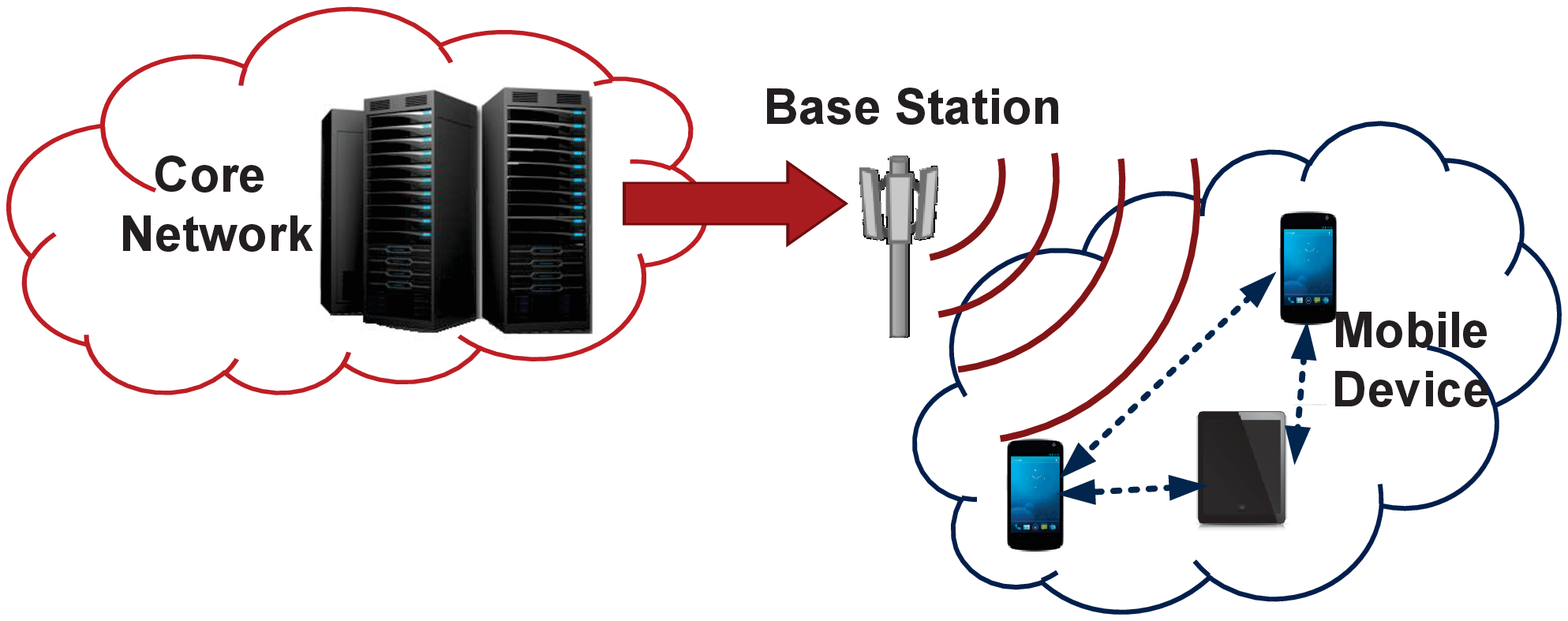}} }
\vspace{-5pt}
\caption{
(a) The default operation in today's cellular systems: Each mobile device receives its data via unicast transmission over a cellular link. (b) Cooperative mobile devices with multiple interfaces: Mobile devices can cooperate and use both cellular and D2D interfaces concurrently to efficiently utilize available resources.
}
\vspace{-15pt}
\label{fig:intro_example}
\end{figure}

The default operation in today's networks is to connect each mobile device to the Internet via its cellular or Wi-Fi connection, Fig.~\ref{fig:intro_example}(a). On the other hand, utilizing device-to-device (D2D) connections among mobile devices simultaneously with the cellular connections is promising to meet the increasing throughput demand \cite{MI1, MI2, MI3, MI4, MI5, MI7, microcast, microcast_allerton}. In particular, when mobile devices are in the close proximity of each other and are interested in the same content, D2D connections such as Wi-Fi Direct can be opportunistically used to construct a cooperative (and jointly operating) cellular and D2D networking system, Fig.~\ref{fig:intro_example}(b). 

\begin{figure}[t!]
\begin{center}
\subfigure[Broadcasting four packets; $p_1$, $p_2$, $p_3$, $p_4$]{ \scalebox{.5}{\includegraphics{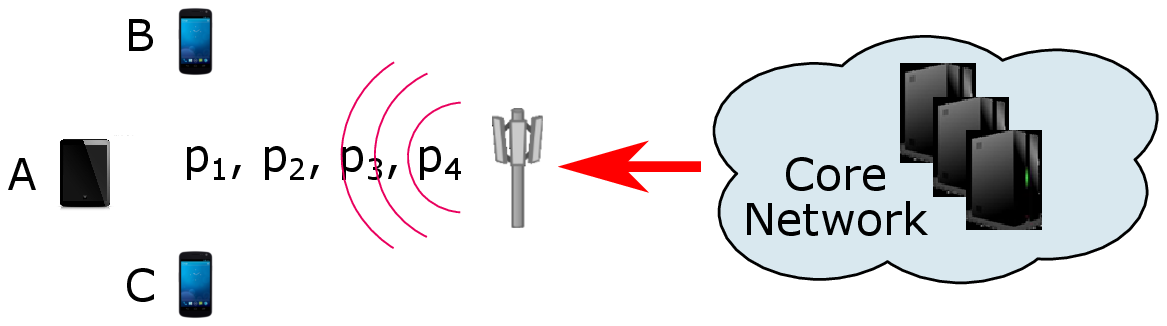}} } %
\subfigure[Missing packets after broadcast]{ \scalebox{.5}{\includegraphics{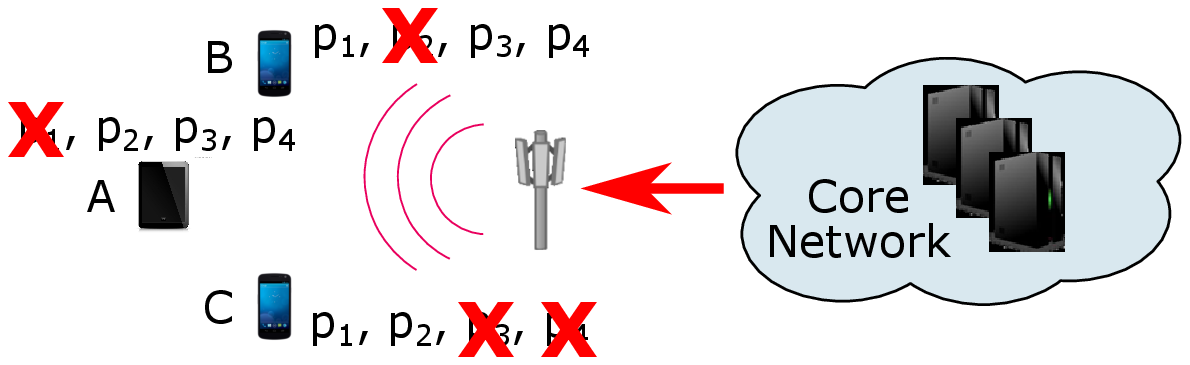}} }
\end{center}
\begin{center}
\vspace{-5pt}
\caption{\label{fig:missing} 
Example scenario with packet losses. (a) Packets $p_1$, $p_2$, $p_3$, and $p_4$ are broadcast from the base station. (b) After the broadcast, $p_1$ is missing at mobile device $A$, $p_2$ is missing at $B$, and $p_3$ and $p_4$ are missing at $C$. }
\vspace{-20pt}
\end{center}
\end{figure}

In this paper, our goal is to understand the potential of the system when D2D networking meets cellular, and develop a network coding framework to exploit this potential. We consider a scenario that a group of mobile devices are in the same transmission range and thus can hear each other. These cooperative mobile devices, that are interested in the same content, \eg video, exploit both cellular and D2D connections. In this setup, a common content is broadcast over cellular links\footnote{\scriptsize Note that broadcasting over cellular links is part of LTE \cite{3gpp_lte_broadcast,ericsson_lte_broadcast,qualcomm_lte_broadcast}, and getting increasing interest in practice, so we consider broadcast scenario instead of unicast.}, Fig. \ref{fig:missing}(a). However, mobile devices may receive only a partial content due to packet losses over cellular links, Fig. \ref{fig:missing}(b). The remaining missing content can then be recovered by utilizing both cellular and D2D connections simultaneously in a cooperative manner. In this setup, thanks to using different parts of the spectrum, cellular links and D2D connections (namely, we consider Wi-Fi Direct) operate concurrently. Thus, a mobile device can receive two packets simultaneously; one via cellular, and the other via D2D connections. The fundamental question in this context, and the focus of this paper, is to design and analysis of efficient network coding algorithms that take into account (i) concurrent operation of cellular and D2D connections, and (ii) the cooperative nature of mobile devices.

The performance of network coding in cellular only and D2D only systems has been considered in previous work, \cite{cope,cc_wang,salim_broadcast,RouayhebITW10,RouayhebISIT10,SprintsonQShine10,TajbakhshAusCTW14,parastoo_broadcast}, in the context of broadcasting a common content over cellular links, and repairing the missing content via retransmissions over cellular links, or by exploiting D2D connections. The following example demonstrates the potential of network coding in cellular and D2D only systems.

\begin{example} \label{example_indv_xmit}
{\em Cellular only setup:} Let us consider Fig. \ref{fig:missing}(a), where four packets, ${p_1,p_2,p_3,p_4}$ are broadcast from the base station. Assume that after the broadcast, $p_1$ is missing at mobile device $A$, $p_2$ is missing at $B$, and $p_3$ and $p_4$ are missing at $C$, Fig. \ref{fig:missing}(b). The missing packets can be recovered via re-transmissions (broadcasts) in a cellular only setup (D2D connections are not used for recovery). Without network coding, four transmissions are required so that each mobile device receives all the packets. With network coding, two transmissions from the base station are sufficient: $p_1+p_2+p_3$ and $p_4$. After these two transmissions, all mobile devices have the complete set of packets. As seen, network coding reduces four transmissions to two, which shows the benefit of network coding in {\em cellular only} setup.

{\em D2D only setup:} Now let us consider packet recovery by exploiting only D2D connections (cellular connections are not used for recovery). Assume that $p_1$ is missing at mobile device $A$, $p_2$ is missing at $B$, and $p_3$ and $p_4$ are missing at $C$. Without network coding, four transmissions are required to recover all missing packets in all mobile devices. With network coding by exploiting D2D connections, two transmissions are sufficient: (i) mobile device $B$ broadcasts $p_1+p_3$, and (ii) $A$ broadcasts $p_2+p_4$. After these two transmissions, all mobile devices have all the packets. In this example, by taking advantage of network coding, the number of transmissions are reduced from four to two transmissions.
\hfill $\Box$
\end{example}

The above example demonstrates the benefit of network coding in {\em cellular only} and {\em D2D only} setups. Yet, mobile devices are not limited to operate in {\em cellular only} and {\em D2D only} setups. Indeed, mobile devices can exploit multiple interfaces simultaneously including cellular and D2D connections. The following example demonstrates the potential of network coding in the {\em joint cellular and D2D} setup.

\begin{example} {\em Joint cellular and D2D setup:} Let us consider Fig. \ref{fig:missing}(b) again, and assume that after the broadcast, $p_1$ is missing at $A$, $p_2$ is missing at $B$, and $p_3$ and $p_4$ are missing at $C$. In order to recover the missing packets, we exploit both cellular and D2D links; each mobile device can receive two simultaneous packets, one of which is transmitted through the cellular links from the source and the other is transmitted through a D2D link from one of the mobile devices. We assume that cellular and D2D links have the same capacity. For this example, the following transmissions are simultaneously made to recover the missing packets: (i) the base station broadcasts $p_1+p_3$ via cellular links, and (ii) mobile device $A$ broadcasts $p_2+p_4$ via D2D links. As seen, the number of transmission slots is reduced to one from two as compared to Example~\ref{example_indv_xmit}.
\hfill $\Box$
\end{example}

As seen, mobile devices that use their cellular and D2D connections simultaneously and cooperatively have a potential of improving throughput significantly. However, it is crucial to understand and quantify the potential of network coding for cooperating mobile devices in joint cellular and D2D setup. In this paper, we consider this problem, and (i) develop a network coding framework, called {\em network coding for multiple interfaces (i.e., jointly operating cellular and D2D interfaces) (NCMI)}, for cooperative mobile devices, where cellular and D2D link capacities are the same, and (ii) characterize the performance of the proposed network coding framework, where we use packet completion time, which is the number of transmission slots to recover all packets, as a performance metric. The following are the key contributions of this work:
\begin{itemize}
\item We propose a network coding algorithm; \ncrn, where packets are network coded as a batch (of packets) to improve the throughput of cooperative mobile devices in a joint cellular and D2D setup. By taking into account the number of packets that each mobile device would like to receive for packet recovery, we develop an upper bound on the packet completion time of \ncrn.
\item For the same joint cellular and D2D setup, we develop a network coding algorithm; \ncin, where packets are network coded in a way that they can be decoded immediately after they are received. \ncin ~is crucial for applications with deadline constraints. Furthermore, we characterize the performance of \ncin, and develop an upper bound on its packet completion time.
\item We develop a lower bound on the packet completion time when any network coding algorithm is employed in the joint cellular and D2D setup. 
\item We evaluate \ncrn~ and \ncin~ for different numbers of devices and packets, and different loss probabilities. The simulation results show that our algorithms significantly improve packet completion time as compared to baselines, and the upper bounds we developed for \ncrn~ and \ncin~ are tight.
\end{itemize}

The structure of the rest of this paper is as follows. Section \ref{sec:preliminaries} presents preliminaries and our problem statement. Section \ref{sec:upper} presents our proposed network coding algorithm, and provides our upper bound analysis. Section \ref{sec:lower} presents the lower bound on the performance of any network coding algorithm. Section \ref{sec:simulation} presents simulation results. Section \ref{sec:related} presents related work. Section \ref{sec:conclusion} concludes the paper.

\vspace{-10pt}
\section{\label{sec:preliminaries}Preliminaries \&  Problem Statement}
\vspace{-5pt}
We consider a setup with $N$ cooperative mobile devices, where $\Nset$ is the set of devices in our system with $N=|\Nset|$. These devices are within close proximity of each other, so they are in the same transmission range. The mobile devices in $\Nset$ are interested in receiving packets $p_m$ from set $\Mset$, \ie $p_m \in \Mset$ and $M = |\Mset|$.

The mobile devices in our system model are able to use cellular and D2D interfaces simultaneously to receive data. In particular, we consider a two-stage model for our joint cellular and D2D setup. In the first stage, all packets are broadcast to all devices via cellular links, while in the second stage, missing packets are recovered by utilizing cellular and D2D links jointly. This two-stage model fits well for error correction, because the cellular only operation in the first stage saves energy (as keeping multiple interfaces open increases energy consumption), and the joint cellular and D2D operation in the second stage helps quickly recover missing packets while relieving the load on cellular links. In our setup, D2D links are only needed if some packets are lost over the cellular broadcast link. Thus, at time slots when there are no packet losses, which is common scenario for practical loss rates, D2D interfaces of devices would remain idle if we keep both cellular and D2D interfaces open in the first stage. However, idle interfaces, \ie even if no packet is transmitted or received, still consume energy \cite{energy1}, \cite{energy2}. Thus, our two-stage model is a good approach for mobile devices operating on batteries. Next, we further explain the operation of our two-stage model.

In the first stage, all packets are broadcast to all devices via cellular links. During the first stage, mobile devices may receive partial content due to packet losses over the cellular broadcast link.  Thus, after the first stage, the set of packets that mobile device $n \in \Nset$ has successfully received is $\Hset_n$, and is referred to as {\em Has} set of device $n$. The set of packets that is lost in the first stage at mobile device $n$ is referred to as {\em Wants} set of device $n$ and denoted by $\Wset_n$; $\Wset_n = \Mset \setminus \Hset_n$. Furthermore, we define the set $\Mset_c$ as $\Mset_c=\bigcap_{n \in \Nset} \Wset_n$. Note that the packets in $\Mset_c$ are not received by any devices during the first stage.

In the second stage, missing packets are recovered by utilizing  cellular and D2D links jointly. In particular, a mobile device may receive two recovery packets; one from cellular link and another from D2D link, simultaneously. Exploiting joint cellular and D2D links has the potential of improving throughput \cite{microcast_allerton}. In order to use the available resources more efficiently, we need to determine the best possible network coded packets to be transmitted over cellular and D2D links at each transmission slot. This is an open problem and the focus of this paper.

In particular, in this paper, we develop a network coding framework for multiple interfaces (NCMI) in a joint cellular and D2D setup to recover missing packets in the second stage. Namely, we develop \ncrn ~based on batch network coding and \ncin ~based on instantly decodable network coding (IDNC) \cite{sorourICC}. In \ncrn, each transmitted packet to device $n \in \Nset$ is a linear combination of the missing packets in that device and thus it carries information about all missing packets. Therefore, all missing packets can be decoded at device $n$ once enough number of packets are received successfully by device $n$. In \ncin, each transmitted packet to device $n \in \Nset$ carries information about only one of the missing packets at that device. Therefore, one missing packet can be decoded each time the transmitted packet is received successfully by device $n$.

The integral part of our work is to analyze the throughput performance of \ncrn ~and \ncin. The amount of time required to recover the missing packets is an indicator of resource (such as energy, time, and bandwidth) consumption. Therefore, we consider the packet completion time as the performance metric, which is defined as follows:
\begin{definition}
\label{def:completion}
Packet completion time $T$ is the number of transmission slots in the second stage that are required for all mobile devices to receive and decode all packets in their {\em Wants} sets.
\end{definition}

In our joint cellular and D2D setup, $\eta_n, n \in \Nset$ denotes the loss probability over the cellular link towards device $n$ and $\epsilon_{k,l}$ denotes the loss probability over the D2D link when device $k$ transmits a packet to device $l$. We assume that $\eta_n$ and $\epsilon_{k,l}$ are i.i.d. according to a uniform distribution.

{\em Assumptions:} We assume, without loss of generality, that for each packet $p_m \in \Mset$, there is at least one mobile device that wants packet $p_m$. In other words, $\forall p_m \in \Mset, \exists n \in \Nset$ such that $p_m \in \Wset_n$. This assumption does not violate generality, because packets that are not wanted by any of the devices could be removed from $\Mset$.

\vspace{-5pt}
\section{\label{sec:upper} NCMI and Upper Bounds on $T$}
\vspace{-5pt}
In this section, we develop network coding algorithms with multiple interfaces (NCMI) for the joint cellular and D2D setup, and provide upper bounds on their packet completion time. In particular, we develop two network coding algorithms; \ncrn, which uses random linear network coding, and \ncin, which provides instant decodability guarantee. We first consider the case of no loss in the second stage, while there is loss in the first stage, and analyze \ncrn ~and \ncin. The analysis with no loss provides us insight while designing network coding algorithms for the lossy scenario in the second stage.\footnote{We also note that computational complexity and signaling overhead of lossless NCMI’s are lower as compared to their lossy versions as we will demonstrate later in this section and Appendix F. Thus, if D2D links are lossless, we can directly use lossless NCMI to enjoy lower computational complexity and signaling overhead.}

\vspace{-5pt}
\subsection{No Loss in the Second stage}
In this section, we assume that cellular connections in the first stage are lossy, but both the cellular and D2D connections are lossless in the second stage. Thus, all the transmitted packets in the second stage are received correctly. We develop network coding algorithms \ncrn ~and \ncin, and develop upper bounds on the packet completion time.

\subsubsection{\ncrn} \label{sec:Batch_lossless}
In this section, we explain and analyze \ncrn.
\paragraph{Algorithm Description}
As we mentioned earlier in Section~\ref{sec:preliminaries}, our system model consists of two stages. In the first stage, all packets are broadcast to all devices via cellular links without network coding. In the second stage, both cellular and D2D links are utilized simultaneously and network coding is employed. In particular, both the source and one of the devices in the local area transmit network coded packets simultaneously at each transmission slot until there is no missing packet in the local area. In \ncrn ~the transmitted packets are formed as a linear combination of the missing packets and thus carry information about all packets in the set $\Mset$. Therefore, to minimize the packet completion time, the network coded packets to be transmitted from the source (through the cellular links) and in the local area among the mobile devices (through D2D links) are selected such that they contain as much information as possible about all missing packets. Next, we explain the details on how network coding is performed by the source and the local area devices at each transmission slot.

The source (i) determines the missing packets in all mobile devices, and (ii) transmits linear combinations of these packets (using random linear network coding over a sufficiently large field) through cellular links. These network coded packets are innovative and beneficial for any device $n$ for which $|\Hset_n| \le M$, because these network coded packets carry information about all missing packets in the local area. After each transmission, if the received packet is innovative for device $n$, it is inserted into $\Hset_n$ set. The procedure continues until each device $n$ receives $|\Wset_n|$ innovative packets.

On the other hand, in the local area, the mobile device $n_{\max}$ with the largest {\em Has} set has the most information about the missing packets among all cooperating devices. Therefore, at each transmission slot, one of the devices is selected randomly as the controller; the controller selects $n_{\max}=\arg \max_{n \in \Nset} |\Hset_n|$ as the transmitter. If there are multiple of such devices, one of them is selected randomly. The transmitter linearly combines all packets in its {\em Has} set, $\Hset_{n_{\max}}$, and broadcasts the network coded packet to all other mobile devices via D2D links. This network coded packet is beneficial to any device $k \neq n_{max}$ (any device except for the transmitter) as long as $\Wset_k \setminus \Mset_c$ is not an empty set. The reason is that device $n_{max}$ has the most number of packets from the set $\Mset$ among all devices. Therefore, $n_{max}$ is the only device that has innovative information about the missing packets at all devices $k \in (\Nset \setminus n_{max})$.
After each transmission, if the received packet has innovative information for device $n$, the received packet is inserted into the {\em Has} set of device $n$. Note that the network coded packets that include packets from $\Mset_c=\bigcap_{n \in \Nset} \Wset_n$ can only be transmitted from the source, since these packets do not exist in any of the devices. Therefore, the devices stop transmitting network coded packets through D2D links if each device $n$ receives (i) $|\Wset_n|-|\Mset_c|$ innovative packets from D2D links, or (ii) $|\Wset_n|$ innovative packets from both the cellular and D2D links.

Next, we characterize how long it takes until all missing packets are recovered and calculate the packet completion time, $T$.

\paragraph{Upper Bound on $T$}
In order to characterize the performance of proposed \ncrn, we consider the worst case scenario and develop an upper bound on the packet completion time of \ncrn. We note that  \ncrn ~is guaranteed to outperform the upper bound, as proved in Appendix A and shown in the simulation results.
\begin{theorem} \label{th:easy_upper_bound}
Packet completion time; $T$ when \ncrn ~is employed by cooperative mobile devices on a joint cellular and D2D setup is upper bounded by
\begin{equation} \label{eq:b2_random}
T \leq \left \lceil \max\Big(|\Mset_c|,\frac{1}{3}\big(\max_{n \in \Nset} |\Wset_n|+\min_{n \in \Nset} |\Wset_n|\big), \frac{1}{2} \max_{n \in \Nset} |\Wset_n|\Big) \right \rceil.
\end{equation}
\end{theorem}
{\em Proof:} The proof is provided in Appendix A.\hfill $\blacksquare$

\begin{example}
Let us consider three mobile devices with the {\em Wants} sets; $\Wset_A=\{p_1,p_2,p_3\}, \Wset_B=\{p_1,p_4,p_5\}, \Wset_C=\{p_1,p_6,p_7\}$. Using \ncrn, in the first transmission slot, the source transmits a linear combination of the packets $p_1, p_2, ..., p_7$, which is innovative for all devices $A$, $B$, and $C$, as it carries information about all missing packets. Meanwhile, in the local area, the device with the largest {\em Has} set is selected. Since there is equality in this example ($|\Hset_A|=|\Hset_B|=|\Hset_C|=4$), one device is selected randomly, let us say device $A$. Device $A$ transmits a linear combinations of $p_4, p_5, p_6, p_7$ via D2D links. Note that this network coded packet is beneficial to both devices $B$ and $C$, as it carries information about their missing packets. Therefore, at the end of the first transmission slot, device $A$ receives one innovative packet and thus the size of its {\em Has} set is increased by one and devices $B$ and $C$ receive two innovative packets and thus the sizes of the {\em Has} sets for these devices are increased by two; \ie $|\Hset_A|=5$, $|\Hset_B|=|\Hset_C|=6$. In the second transmission slot, the source transmits a linear combination of $p_1, p_2, ..., p_7$, which is innovative for all devices $A$, $B$, and $C$ and at the same time $B$ or $C$ (with the larger size of {\em Has} set than $A$) transmits a linear combination of the packets in its {\em Has} set, which is innovative for $A$. Therefore, at the end of the second transmission slot, device $A$ receives two innovative packets and thus the size of its {\em Has} set is increased by two and devices $B$ and $C$ receive one innovative packet (they only need one innovative packet to be satisfied) and thus the sizes of the {\em Has} sets for these devices are increased by one; \ie $|\Hset_A|=7$, $|\Hset_B|=|\Hset_C|=7$. As seen, each device receives three innovative packets after two transmission slots and thus the packet completion time is equal to $T=2$. On the other hand, from Theorem \ref{th:easy_upper_bound}, the upper bound for the packet completion time is equal to $\left \lceil \max\big(1,\frac{1}{3}(3+3), \frac{3}{2}\big) \right \rceil = 2$. As seen, the inequality $T \leq 2$ (Eq. \ref{eq:b2_random}) is satisfied, in this example. It can also be seen that the upper bound is tight, in this example.
\end{example}

\paragraph{Computational Complexity}
In \ncrn, at each transmission slot, the source creates a linear combination of all packets with the complexity of $O(M)$. Meanwhile, among all devices, the one with the maximum size of {\em Has} set is selected as the transmitter with the complexity of $O(N)$. Then, the transmitter creates a linear combination of the packets in its {\em Has} with the complexity of $O(M)$. Therefore, the computational complexity at each transmission slot is $O(M+N+M)=O(M+N)$. As the maximum number of transmission slots is equal to the number of missing packet, $M$, the complexity of \ncrn ~(when the channels are lossless in the second stage), is polynomial with the complexity of $O(M(M+N))=O(M^2+N)$.

\subsubsection{\ncin} \label{sec:ncin_no_loss}
In this section, we develop and analyze \ncin, which is a heuristic to network code packets in a way that they can be decoded immediately after they are received by the mobile devices. \ncin ~is crucial for loss tolerant real-time applications with deadline constraints.

\paragraph{Algorithm Description} \label{sec:ncin_algorithm}
\ncin ~determines the IDNC packets to be transmitted at each transmission slot through the cellular and D2D links with the goal of minimizing the packet completion time. To reach this goal, the optimum way is exhaustively creating all combinations of IDNC packets that can be transmitted from the source and the mobile devices in all transmission slots and selecting the sequence of packets that results in the minimum average packet completion time. However, the complexity of exhaustive search is high and thus in this paper, we proposed a heuristic method \ncin ~with linear complexity with respect to the number of devices and quadratic complexity with respect to the number of packets. \ncin ~consists of three steps: (i) creating IDNC packets (Algorithm 1), (ii) grouping the created IDNC packets into the sets $\Mset_c$, $\Mset_l$ and $\Mset_d$ (Algorithm 1), and (iii) determining the IDNC packets to be transmitted from the two interfaces based on the the created groups. In the following, we explain these steps.

\begin{algorithm}[h]
\caption{Grouping the packets in the {\em Wants} Sets}
\label{al:Groups}
\begin{algorithmic}[1]
\FOR{any packet $p_m$ in $\Mset$}
\STATE Define vector $\boldsymbol v_m$ with size $N$. Each element of the vector $\boldsymbol v_m$ is initially set to $NULL$; \ie $\boldsymbol v_m[n] = NULL$, $\forall n \in \Nset$.
\FOR{any device $n$ in $\Nset$}
\IF {$p_m$ is wanted by device $n$}
\STATE $\boldsymbol v_m[n] = p_m$ and $p_m$ is removed from the {\em Wants} set $\Wset_n$.
\ENDIF
\ENDFOR
\IF {there exists a vector $\boldsymbol v_{m'}$, $m' < m$ satisfying for any $n$, for which $\boldsymbol v_{m}[n] = p_m$ then $\boldsymbol v_{m'}[n] = NULL$}
\STATE Replace $\boldsymbol v_{m'}$ with $\boldsymbol v_{m'} + \boldsymbol v_m$ and delete $\boldsymbol v_{m}$. (Note that $p_m + NULL = p_m$ for any $m$)
\ENDIF
\ENDFOR
\STATE Each element of $\Mset_c$ is constructed by network coding all packets in a vector $v_m$ if all elements of $v_m$ are the same and not equal to $NULL$; \ie $v_m[1] \neq NULL$ and $v_m[n]=v_m[1]$, $\forall n \in \Nset$.
\STATE
Each element of $\Mset_d$ is constructed by network coding all packets in a vector $\boldsymbol v_m$ if $v_m$ contains at least one element equal to $NULL$; \ie $\exists x \in \Nset \mid v_m[x]=NULL$.
\STATE Construct $\Mset_l$ using the remaining vectors. In other words, each element of $\Mset_l$ is constructed by network coding all packets in a vector $v_m$ if $v_m$ does not contain any $NULL$ element and contains at least two different elements; \ie $v_m[n] \neq NULL, \forall n \in \Nset$ and $\exists n, x \mid v_m[n] \neq v_m[x]$.
\end{algorithmic}
\end{algorithm}

{\bf Step 1: Creating IDNC packets:} Algorithm 1 is a greedy algorithm that creates IDNC packets, sequentially, from the packets in the set of missing packets in all devices. The main idea behind this algorithm is to check uncoded packets sequentially and try to merge them into IDNC packets. We note that similar ideas have been considered in network coding literature, \eg a greedy algorithm for creating network coding packets over wireless mesh networks is developed using a similar approach in \cite{cope}. According to Algorithm 1, the first IDNC packet is created with the first uncoded packet. Then for each remaining uncoded packets, the algorithm checks all of the previously created IDNC packets; if the uncoded packet can be merged with the IDNC packet (\ie the merged IDNC packet is instantly decodable for all devices that want one of the uncoded packets in the IDNC packet), the IDNC packet would be updated by adding the uncoded packet. Otherwise, a new IDNC packet containing the uncoded packet is created. This algorithm creates IDNC packets with complexity of $O(M^2+N)$. Note that the complexity of exhaustive search to create all possible IDNC packets is NP-hard. Next, we describe the details of Algorithm 1 on Step 1, creating IDNC packets.

In Algorithm \ref{al:Groups}, we define vectors with length $N$, whose elements are either $NULL$ or $p_m \in \Mset$. A network coded packet is associated to each vector and constructed as a linear combination of all elements of the vector that are not equal to $NULL$. We describe how each vector is defined by Algorithm \ref{al:Groups} in the following. First, the vector $\boldsymbol v_{m}$ with the length of $N$ is defined for each packet $p_m \in \Mset$. The $n$th element of this vector, where $n$ is any device that wants that packet, is initially set to $p_m$. The value of this vector for the remaining elements, which correspond to the devices that have that packet, is initially set to $NULL$ (lines 1-5). A network coded packet is instantly decodable for device $n$ if it contains information about one and only one of the packets in the {\em Wants} set of device $n$. Let us consider packet $p_m$ with its corresponding vector $\boldsymbol v_{m}$ and the instantly decodable network coded packet corresponding to vector $\boldsymbol v_{m'}$. These two packets can be combined and merged as a new instantly decodable network coding packet (line 7) if the value of vector $\boldsymbol v_{m'}$ for all devices that want packet $p_m$ (\ie any device $n$ for which $\boldsymbol v_m[n]$ is $p_m$) is $NULL$ (the condition of {\em if} statement at line 6).

{\bf Step 2: Grouping the created IDNC packets:} After creating the IDNC packets, we group them into the sets, $\Mset_c$, $\Mset_l$, and $\Mset_d$, based on their differences from the view point of cellular and D2D links. The packets in $\Mset_c$ can only be transmitted via cellular links. The packets in $\Mset_l$ can be transmitted via  cellular or D2D links, but it may take more time slots if they are transmitted via D2D links. The packets in $\Mset_d$ can be transmitted either via cellular or D2D links, and the number of time slots for transmitting a packet is the same in both cellular and D2D. The sets $\Mset_c$, $\Mset_l$ and $\Mset_d$ are created as the following. $\Mset_c$ is the set of packets that are not available in the local area and should only be re-broadcast from the base station (source) to all devices. These are the packets that are wanted by all devices. $\Mset_d$ is the set of IDNC packets that can be transmitted by a single transmission from either the source or a mobile device. These are the packets that can be formed by combining a subset of the packets in the {\em Has} set of one of the devices (note that any combination of packets can be created in the source as it has all packets). $\Mset_l$ is the set of remaining IDNC packets that can be transmitted by a single transmission from the source or by two transmissions (according to Lemma \ref{th:M_l_set}) in the local area. These are the packets that contain one and only one uncoded packet from the {\em Wants} set of each device and thus they are not available in one single device. Therefore, in order to send these packets through D2D links, they need to be divided into two IDNC packets and each part can be transmitted from the device that has all the corresponding uncoded packets.

\begin{lemma} \label{th:M_l_set}
Each network coded packet in $\Mset_l$ can be transmitted by exactly two transmission slots using D2D links by splitting the packet into two (network coded) packets.
\end{lemma}

{\em Proof:} Each network coded packet in $\Mset_l$ contains one and only one packet from each $\Wset_n, \forall n \in \Nset$ set. Therefore, there is no single device among the cooperating devices that has all uncoded packets of the network coded packet, and thus one transmission slot is not sufficient to transmit the network coded packet in the local area. On the other hand, since there are at least two different uncoded packets in the network coded packet, we can split the network coded packet into two (network coded) packets; each device wants one uncoded packet from one of the splitted packets and has all the uncoded packets in the other splitted packet. Therefore, each of the two splitted packets can be transmitted from at least one device among the cooperating devices and thus two transmissions are necessary and sufficient to transmit the content of the network coded packet via D2D links.
\hfill $\blacksquare$

The properties of the sets $\Mset_c, \Mset_l,$ and $\Mset_d$ are summarized in Table~\ref{table:diff}.

\begin{table}
\caption{The required number of transmissions to transmit each packet in $\Mset_c, \Mset_l,$ and $\Mset_d$ via cellular and D2D links.}
\centering
\begin{tabular}{c c c }
\hline
Set & Cellular Link & D2D Link \\ [0.5ex]
\hline 
$\Mset_c$ & $1$ & N/A \\ 
$\Mset_l$ & $1$ & $2$ \\
$\Mset_d$ & $1$ & $1$ \\
\end{tabular} \label{table:diff}
\end{table}

Next, we describe the details of Algorithm 1 on Step 2, grouping the created IDNC packets.

The set $\Mset_c$ consists of all packets that are wanted by all devices; \ie by using any vector $\boldsymbol v_m$ whose elements are the same and not equal to $NULL$ (line 8). The set $\Mset_d$ consists of the network coded packets that can be formed by combining a subset of the packets in the {\em Has} set of one of the devices, called $x$; \ie by using the packets in the vector $\boldsymbol v_m$ whose $x$th element is $NULL$ (line 9). The set $\Mset_l$ consists of the rest of network coded packets. In other words, each network coded packet in $\Mset_l$ contains one and only one packet from the {\em Wants} set of each device; \ie by using any vector $\boldsymbol v_m$ whose elements are not equal to $NULL$ and it contains at least two different elements (line 10).

Next, we give an example on how Algorithm \ref{al:Groups} works.

\begin{example}\label{ex:group}
Let us assume that there are three mobile devices with the {\em Wants} sets; $\Wset_A=\{p_1,p_4\}$, $\Wset_B=\{p_1,p_2,p_3\}$, $\Wset_C=\{p_1,p_4,p_5\}$. Note that $\Mset= \bigcup_{n \in \Nset} \Wset_n = \{p_1,p_2,p_3,p_4,p_5\}$ and $\Hset_n = \Mset \setminus \Wset_n$ for $n \in \{A,B,C\}$; $\Hset_A=\{p_2,p_3,p_5\}$, $\Hset_B=\{p_4,p_5\}$, $\Hset_C=\{p_2,p_3\}$.
According to Algorithm~\ref{al:Groups}, $\boldsymbol v_{1}$ is equal to $[p_1,p_1,p_1]$, because it is wanted by all three devices, as shown in Table \ref{table:ex_group}. Then, for packet $p_2$ we define vector $\boldsymbol v_{2}=[NULL,p_2,NULL]$, because $p_2$ is wanted by device $B$ only. Similarly, for packet $p_3$, we define vector $\boldsymbol v_{3}=[NULL,p_3,NULL]$. $\boldsymbol v_3$ can not be merged with $\boldsymbol v_{1}$ or $\boldsymbol v_{2}$, because the second element of $\boldsymbol v_{3}$ is $p_3$, while the second elements of $\boldsymbol v_{1}$ and $\boldsymbol v_{2}$ are not $NULL$. In the next step, we define vector $\boldsymbol v_{4}=[p_4,NULL,p_4]$. $\boldsymbol v_{4}$ can be merged with $\boldsymbol v_{2}$, because the first and third elements of $\boldsymbol v_{4}$ is $p_4$ and the first and the second elements of $\boldsymbol v_{2}$ is $NULL$. Therefore, $\boldsymbol v_2$ is updated as $\boldsymbol v_{2}+\boldsymbol v_{4}=[p_4,p_2,p_4]$ and $\boldsymbol v_{4}$ is deleted, as shown in Table \ref{table:ex_group}. Similarly, $\boldsymbol v_5$ is defined as $[NULL,NULL,p_5]$. $\boldsymbol v_{5}$ can be merged with $\boldsymbol v_{3}$, because the third element of $\boldsymbol v_{5}$ is $p_5$ and the third element of $\boldsymbol v_{3}$ is $NULL$. Therefore, $\boldsymbol v_{3}$ is updated as $\boldsymbol v_{3}+\boldsymbol v_{5}=[NULL,p_3,p_5]$ and $\boldsymbol v_{5}$ is deleted, as shown in Table \ref{table:ex_group}.

All elements of $\boldsymbol v_{1}$ are the same and not equal to $NULL$. Therefore, $\Mset_c$ is constructed from $\boldsymbol v_1$; $\Mset_c=\{p_1\}$, as shown in Table \ref{table:ex_group}. $\boldsymbol v_3$ has a $NULL$ element. Therefore, $\Mset_d$ is constructed from $\boldsymbol v_3$; $\Mset_d=\{p_3 + p_5\}$. $\boldsymbol v_2$ does not have any $NULL$ element and its first and second elements are different. Therefore, $\Mset_l$ is constructed from vector $\boldsymbol v_2$; $\Mset_l=\{p_2 + p_4\}$.

\begin{table}
\caption{Grouping the packets in the {\em Wants} sets for Example \ref{ex:group}.}
\centering
\begin{tabular}{c| c c c c}
\hline
Set & & $\Wset_A$ & $\Wset_B$ & $\Wset_C$ \\ [0.5ex]
\hline
$\Mset_c$ & $\boldsymbol v_1:$  & $p_1$ & $p_1$ & $p_1$ \\ 
$\Mset_l$ & $\boldsymbol v_2:$ & $p_4$ & $p_2$ & $p_4$ \\
$\Mset_d$ & $\boldsymbol v_3:$ & $NULL$ & $p_3$ & $p_5$ \\
\end{tabular} \label{table:ex_group}
\end{table}

\hfill $\Box$
\end{example}

{\bf Step 3: Determining the IDNC packets to be transmitted from the two interfaces:} At each transmission slot, two packets are selected from the sets $\Mset_c \cup \Mset_l \cup \Mset_d$; one to be transmitted from the base station and another one to be transmitted from one of the mobile devices. The idea behind packet selection in lossless \ncin ~is that the created IDNC packets can be transmitted in the minimum packet completion time.
Note that the decisions of which network coded packet is transmitted and which device is selected as the transmitter in the local area, are made by the controller (which can be selected randomly among mobile devices). {\bf Packet Selection by the Source:} The packets in $\Mset_c$ should only be transmitted from the source. On the other hand, a packet transmission from the set $\Mset_c$ targets all devices. Therefore, the source selects a packet from $\Mset_c$ to transmit, if this set is not empty. Among the sets $\Mset_l$ and $\Mset_d$, the packets in $\Mset_l$ targets all devices, while the packets in $\Mset_d$ targets a subset of devices. On the other hand, packets in $\Mset_l$ require less number of transmissions if transmitted from the source than the mobile devices. Therefore, the source selects a packet from $\Mset_l$ to transmit if $\Mset_c$ is empty. At last, if both $\Mset_c$ and $\Mset_l$ are empty, a packet in $\Mset_d$, is selected to be transmitted from the source. {\bf Packet Selection by the Mobile Devices:} Any packet from the set $\Mset_d$ can be transmitted from the local area by a single transmission, however only a partial of a packet from the set $\Mset_l$ can be transmitted by a single transmission from the local area. Therefore, the order of transmitting the packets in the local area is (i) $\Mset_d$ and (ii) $\Mset_l$.

\paragraph{Upper Bound on $T$}
Now, we analyze the packet completion time performance of \ncin ~by developing an upper bound on the packet completion time.

\begin{theorem}\label{theorem:better_upper}
Packet completion time when \ncin ~is employed by cooperative mobile devices on a joint cellular and D2D setup is upper bounded by:
\begin{align} \label{eq:better_upper}
&T \leq \lceil \mathlarger{\min}\Big(\mathlarger{\max} \big(\frac{M}{2},|\Mset_c|\big), \\ \nonumber
&\mathlarger{\max}\big(|\Mset_c|,\frac{1}{3}\big(2 \min_{n \in \Nset} |\Wset_n|+|\Mset_d|), \frac{1}{2}(\min_{n \in \Nset} |\Wset_n|+|\Mset_d|)\big)\Big) \rceil.
\end{align}
\end{theorem}
{\em Proof:} The proof is provided in Appendix B. \hfill $\blacksquare$

\begin{example}\label{ex:ncin}
Let us assume that there are three mobile devices with the {\em Wants} sets; $\Wset_A=\{p_1,p_2,p_4,p_7,p_9\}$, $\Wset_B=\{p_1,p_2,p_5,p_7,p_{10}\}$, $\Wset_C=\{p_1,p_3,p_6,p_8\}$. Note that $\Mset= \bigcup_{n \in \Nset} \Wset_n = \{p_1,\ldots,p_{10}\}$ and $\Hset_n = \Mset \setminus \Wset_n$ for $n \in \{A,B,C\}$; $\Hset_A=\{p_3,p_5,p_6,p_8,p_{10}\}$, $\Hset_B=\{p_3,p_4,p_6,p_8,p_9\}$, $\Hset_C=\{p_2,p_4,p_5,p_7,p_9,p_{10}\}$.
According to Algorithm~\ref{al:Groups}, $\Mset_c=\{p_1\}$, $\Mset_d=\{p_9+p_{10}\}$, and $\Mset_l=\{p_2 + p_3, p_4 + p_5 + p_6, p_7 + p_8\}$ in this example. Using \ncin, in the first transmission slot packet $p_1$ (in set $\Mset_c$) is transmitted from the source and packet $p_9 + p_{10}$ (in set $\Mset_d$) is transmitted from device C. In the second transmission slot, packet $p_2 + p_3$ (in set $\Mset_l$) is transmitted from the source and packet $p_7$ (the first splitted packet of the third network coded packet in set $\Mset_l$ available at device C) is transmitted from device C. In the third transmission slot, packet $p_4+p_5+p_6$ (in set $\Mset_l$) is transmitted from source and packet $p_8$ (the second splitted packet of the third network coded packet in set $\Mset_l$ available at device A and B) is transmitted from device A or B. Therefore, in total three transmission slots are required by \ncin; $T=3$. On the other hand, from Theorem \ref{theorem:better_upper}, the upper bound for the packet completion time is equal to $\lceil \mathlarger{\min} \big(5,\mathlarger{\max}(1,3,2.5)\big) \rceil = 3$. As seen, the inequality $T \leq 3$ (Eq. \ref{eq:better_upper}) is satisfied, in this example. It can also be seen that the upper bound is tight, in this example.

\hfill $\Box$
\end{example}

\paragraph{Computational Complexity}
Algorithm \ref{al:Groups} constructs vectors by checking all packets and devices with complexity of $O(MN)$. Each constructed vector is checked whether it can be merged with the other vectors; since the maximum number of vectors is $M$, the complexity of merging the vectors is $O(M^2)$. In \ncin, at each transmission slot, the source creates a linear combination of all packets in one of the vectors with the complexity of $O(M)$. Meanwhile, in the local area, one vector is chosen, the device that has all the uncoded packets of the vector in its {\em Has} set is selected as the transmitter with the complexity of $O(N)$, and a linear combination of the packets corresponding to the vector is created with the complexity of $O(M)$ to be transmitted from the transmitter device. Therefore, the complexity of creating and transmitting the packets is $O(M+N+M)=O(M+N)$ for one transmission slot and thus $O(M^2+MN)=O(M^2+N)$ (with at most $M$ transmission slots) for all transmission slots. This complexity is added to the complexity of Algorithm \ref{al:Groups}, $O(M^2+N)$. Thus, the complexity of \ncin ~when the channels are lossless in the second stage is polynomial with $O(M^2+N)$. 

\subsection{Lossy Links in the Second Stage}
\begin{figure}[t!]
\centering
\subfigure[
No loss in the second stage]{ \scalebox{.3}{\includegraphics{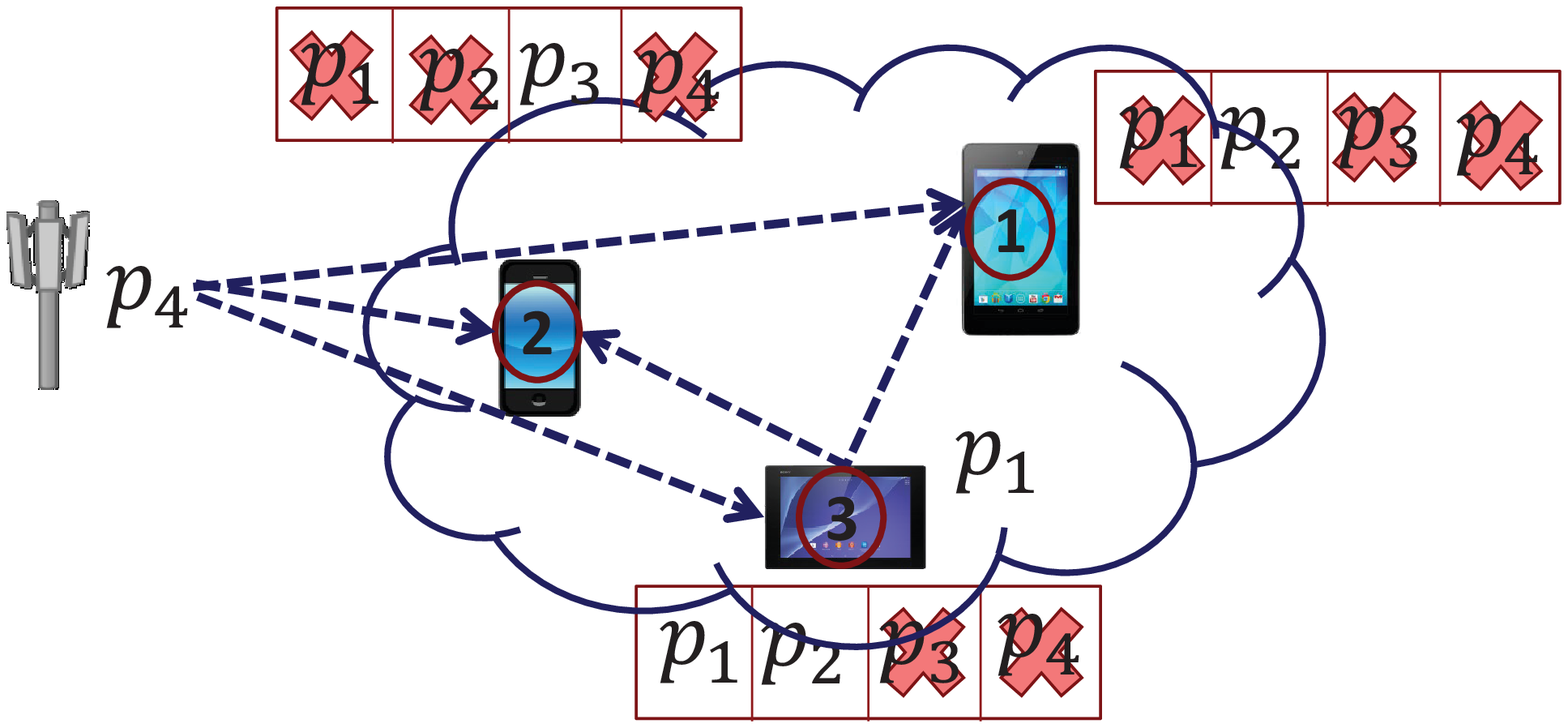}} } \hspace{5pt}
\subfigure[
Loss in the second stage]{ \scalebox{.3}{\includegraphics{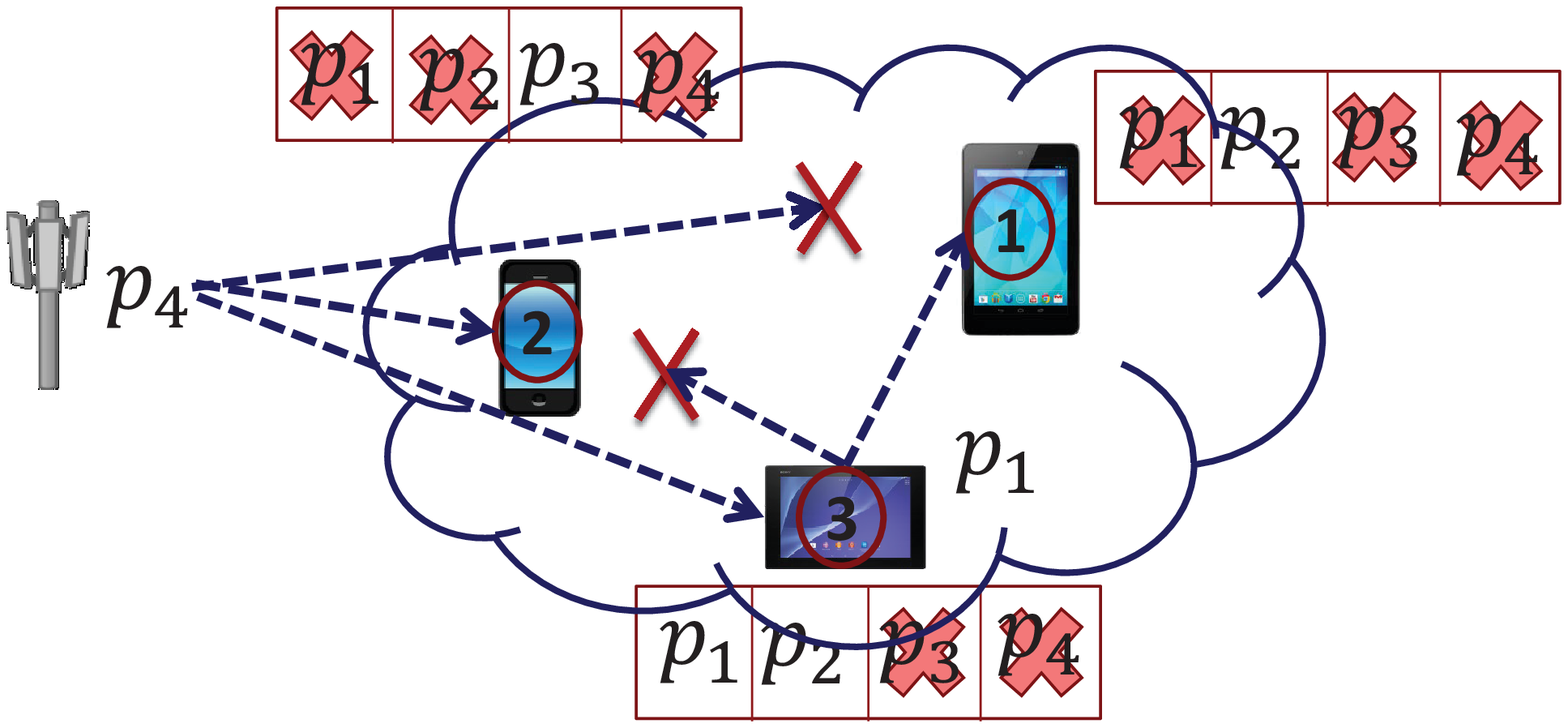}} }
\caption{
To recover the missing packets in the second stage, $p_1$ is selected to be transmitted from device $3$ and $p_4$ is selected to be transmitted from the source. (a) Lossless links in the second stage: All packets are received successfully by the receivers. (b) Lossy links in the second stage: The packet that is transmitted from device $3$ is lost in device $2$ and received successfully in device $1$. Packet transmitted from the source is lost in device $1$ and received successfully in devices $2$ and $3$.
}
\vspace{-13pt}
\label{fig:Loss_sec_phase_ex}
\end{figure}

In the previous section, we developed the network coding algorithms \ncrn ~and \ncin ~for the case that packets in the second stage are received successfully in the receiver devices without any loss (Fig. \ref{fig:Loss_sec_phase_ex}(a)). In this section, we consider a more realistic scenario, where the channels are lossy in the second stage (Fig. \ref{fig:Loss_sec_phase_ex}(b)).

We assume that each packet transmitted from the source to each device $n \in \Nset$ is lost with probability of $\eta_n$. Similarly, each packet transmitted from the transmitter device $k$ to the receiver device $l$ is lost with probability of $\epsilon_{k,l}$. We also assume that probability of all cellular channel losses, $\eta_n, \forall n \in \Nset$, and D2D channel losses, $ \epsilon_{k,l}, \forall k,l \in \Nset$, are i.i.d. and uniformly distributed.

The packet completion time for the lossy links in the second stage, depends on the packets that are received successfully in addition to the transmitted packets at each transmission slot. Therefore, for each transmitted packet, we define {\em targeted receivers}, $\Nset_r$, as the set of devices for which the transmitted packet is innovative. We also define {\em successful receivers} as the set of devices for which the transmitted packet is innovative and is received successfully. We consider the average number of successful receivers for each transmitted packet, which is calculated as follows.

Average number of successful receivers, for a packet transmitted from the source to the set of targeted receivers $\Nset_r$, is equal to the average number of packets that are received successfully at the devices in the set $\Nset_r$ and calculated as $\sum_{n \in \Nset_r} (1-\eta_n)$. Similarly, average number of successful receivers for a packet transmitted from the transmitter device $t$ to the set of targeted receivers $\Nset_r$ is equal to $\sum_{n \in \Nset_r} (1-\epsilon_{t,n})$.

Note that, in the case of no loss, the average number of successful receivers with the set of targeted receivers, $\Nset_r$, is equal to the size of this set; $|\Nset_r|$.

\subsubsection{\ncrn}
In this section, we describe and analyze \ncrn.
\paragraph{Algorithm Description} As described in Section \ref{sec:Batch_lossless}, for the case of lossless channels in the second stage, network coding decisions by \ncrn ~are made such that the transmitted packets contain as much information as possible about missing packets in all devices. For the case of lossy channels, the transmitted packets are selected such that the {\em successfully delivered} packets contain as much information as possible about the missing packets in all devices. In other words, the transmitted packets are selected such that the average number of successful receivers are maximized. Packet selection algorithm by \ncrn ~at each transmission slot for the lossy channels is provided in Algorithm \ref{al:ncrn_loss}. According to this algorithm, two packets are selected; one to be transmitted from the source and the other one to be transmitted in the local area. The details of packet selection in Algorithm \ref{al:ncrn_loss} are described next.

\begin{algorithm}[h]
\caption{\ncrn ~for Lossy Channels}
\label{al:ncrn_loss}
\begin{algorithmic}[1]

\STATE $\Mset$: the set of the missing packets in all devices.

{\bf Packet Selection by the Source:}
\STATE A network coded packet is generated as a linear combination of all packets in $\Mset$.
\STATE The source broadcasts this packet to all mobile devices.

{\bf Packet Selection by the Mobile Devices:}
\FOR{any device $n$ in $\Nset$}
\STATE $\Nset_r=\{k \mid (\exists p \in \Hset_n \mid p \rotatebox[origin=c]{90}{$\models$} \Hset_k)\}$. \Ie packet $p$ (which could be network coded or not) cannot be expressed as a linear combination of the packets in $\Hset_k$.
\STATE $Ave_n=\sum_{k \in \Nset_r} (1-\epsilon_{n,k})$.
\ENDFOR

\STATE $x=\arg\max_{n \in \Nset} Ave_n$.
\STATE A network coded packet is generated as a linear combination of all packets in $\Hset_{x}$.
\STATE $x$ broadcasts this packet to all other mobile devices..

\end{algorithmic}
\end{algorithm}

{\bf Packet Selection by the Source (lines 1-2):} The average number of successful receivers, for a transmitted packet from the source with the set of targeted receivers, $\Nset_r$, is equal to $\sum_{n \in \Nset_r} (1-\eta_n)$. On the other hand, any device $n$ with $|\Wset_n| > 0$ or equivalently $|\Hset_n| < M$, is interested in receiving an innovative packet and can be a member of $\Nset_r$ for the transmitted packet from the source. Therefore, in order to maximize the average number of successful receivers, we need to enlarge the set $\Nset_r$ to all devices that need an innovative packet; \ie $\Nset_r = \{n \mid (|\Hset_n| < M)\}$. According to Algorithm \ref{al:ncrn_loss}, in \ncrn, the source (i) determines the missing packets in all mobile devices, and (ii) transmits linear combinations of these packets (using random linear network coding over a sufficiently large field) as the network coded packet through cellular links. This network coded packet carry information about all missing packets in the local area, and thus is innovative and beneficial for any device $n$ for which $|\Hset_n| < M$; \ie $\Nset_r = \{n \mid (|\Hset_n| < M)\}$. Therefore, the packet selected to be transmitted from the source in \ncrn ~maximizes the average number of successful receivers. After each transmission, if the received packet is innovative for device $n$, it is inserted into $\Hset_n$ set. The procedure continues until each device $n$ receives $|\Wset_n|$ innovative packets.

{\bf Packet Selection by the Mobile Devices (lines 3-7):} In \ncrn, the controller (which can be randomly selected among mobile devices) selects one of the devices as the transmitter (according to line 7). Then, the transmitter transmits a linear combination of the packets in its {\em Has} set to all other devices through D2D links. The set of targeted receivers, $\Nset_r$, for a packet transmitted from device $n$, is the set of devices, for which a random linear combination of the packets in $\Hset_n$ is innovative. In other words, if there is at least one packet from $\Hset_n$ that is linearly independent of all packets in $\Hset_k$, then $k$ is a member of $\Nset_r$; $\Nset_r=\{k \mid (\exists p \in \Hset_n \mid p \rotatebox[origin=c]{90}{$\models$} \Hset_k)\}$\footnote{$\rotatebox[origin=c]{90}{$\models$}$ in $p \rotatebox[origin=c]{90}{$\models$} \Hset_k$ means that packet $p$ (which could be network coded or not) cannot be expressed as a linear combination of the packets in $\Hset_k$.}. Accordingly, the average number of successful receivers for a packet transmitted from $n$ is equal to $\sum_{k \in \Nset_r} (1-\epsilon_{n,k})$. In \ncrn, the mobile device $x$ with the largest average number of successful receivers is selected as the transmitter among all devices at each transmission slot; $x=\arg\max_{n \in \Nset} \sum_{k \in \Nset_r} (1-\epsilon_{n,k})$. If there are multiple of such devices, one of them is selected randomly. The transmitter linearly combines all packets in its {\em Has} set, $\Hset_{x}$, and broadcasts the network coded packet to all other mobile devices. This network coded packet is beneficial to all devices in the set of targeted receivers (by the transmitter) that receive the packet successfully.

After each transmission, if the successfully received packet has innovative information for device $k$, it is inserted into the {\em Has} set of device $k$. Similar to the case of lossless channels, the cooperating devices stop transmitting network coded packets if each device $n$ successfully receives (i) $|\Wset_n| - |\Mset_c|$ innovative packets from the cooperating devices, or (ii) $|\Wset_n|$ innovative packets from both the source and cooperating devices. Note that selecting the device with the largest average number of successful receivers for the case of no loss, $\epsilon_{n,k}=0, \forall n,k \in \Nset$, is equivalent to selecting the device with the largest number of targeted receivers and thus the device with the largest {\em Has} set. This is aligned with our strategy, presented in Section \ref{sec:Batch_lossless}, for packet selection in \ncrn ~when the channels are lossless in the second stage.

Next, we give an example on \ncrn ~for lossy channels.

\begin{example}
Let us consider three mobile devices with the {\em Wants} sets; $\Wset_A=\{p_1,p_2,p_3\}, \Wset_B=\{p_1,p_4,p_5\}, \Wset_C=\{p_1,p_6,p_7\}$, the {\em Has} sets $\Hset_A=\{p_4,p_5,p_6,p_7\}, \Hset_B=\{p_2,p_3,p_6,p_7\}, \Hset_C=\{p_2,p_3,p_4,p_5\}$ and probabilities of channel losses; $\eta_A=0.35, \eta_B=0.4, \eta_C=0.45, \epsilon_{A,B}=0.1, \epsilon_{A,C}=0.3, \epsilon_{B,A}=0.1, \epsilon_{B,C}=0.2, \epsilon_{C,A}=0.3, \epsilon_{C,B}=0.2$. By using \ncrn, $p_1, \ldots, p_7$ are combined as a network coded packet in the source and transmitted to all devices in the first slot. This transmitted packet can be beneficial to all devices ($\Nset_r=\{A,B,C\}$) as it carries information about all missing packets. Note that the average number of successful receivers for this transmitted packet is equal to $(1-0.35)+(1-0.4)+(1-0.45)=1.8$. Meanwhile, in the local area, the device with the largest average number of receivers is selected as the transmitter. The set of targeted receivers for a random linear network coded packet transmitted from $A$ is $\{B,C\}$, as $p_4 \in \Hset_A$ (or $p_5 \in \Hset_A$) is linearly independent of $\Hset_B=\{p_2,p_3,p_6,p_7\}$ and $p_6 \in \Hset_A$ (or $p_7 \in \Hset_A$) is linearly independent of $\Hset_C=\{p_2,p_3,p_4,p_5\}$ and thus the average number of successful receivers is equal to $(1-\epsilon_{A,B})+(1-\epsilon_{A,C})=1.6$. Similarly, the average number of successful receivers for a random linear network coded packet transmitted from $B$ and $C$ is equal to $1.7$ and $1.5$, respectively. Therefore, device $B$ with the maximum average number of successful receivers is selected as the transmitter and transmits a random linear combination of $p_2, p_3, p_6, p_7$. Note that this network coded packet can be received successfully at devices $A$ and $C$ with probabilities of $0.9$ and $0.8$, respectively. Thus, in the first slot, the source transmits linear combination of $p_1, \ldots, p_7$ and device $B$ transmits a random linear combination of $p_2, p_3, p_6,p_7$, simultaneously. The procedure is repeated at every slot until each device receives $3$ innovative packets.
\hfill $\Box$
\end{example}

Next, we characterize how long it takes until all missing packets are recovered; \ie the packet completion time; $T$.

\paragraph{Upper Bound on $T$}
In order to characterize the performance of proposed \ncrn, we provide an upper bound on the packet completion time obtained from \ncrn ~in the following theorem.
\begin{theorem} \label{th:upper_bound_batch_lossy}
The average of packet completion time; $T$ when \ncrn ~is employed by cooperative mobile devices on a joint cellular and D2D setup when the channel links are lossy is upper bounded by
\begin{equation} \label{eq:b2_random_l}
T \leq \left \lceil \max\bigg(\frac{|\Mset_c|}{1-\prod_{n \in \Nset} \eta_{n}},T_j\bigg) \right \rceil,
\end{equation}
\end{theorem}
where,

\begin{align}
T_j=
\begin{cases}
  \max(T_x,T_r), & \mbox{if } \frac{|\Wset_r|-|\Wset_x|}{1-\eta_r-\epsilon_{x,r}+\eta_x} \leq \frac{|\Wset_x|}{1-\eta_x} \\
  \frac{|\Wset_r|}{2-\eta_{r}-\epsilon_{x,r}}, & \mbox{otherwise},
\end{cases}
\end{align}

\begin{equation}\label{eq:Tx}
T_x=\frac{|\Wset_r|(1-\epsilon_{r,x})+|\Wset_x|(1+2\eta_x-2\eta_r+\epsilon_{r,x}-2\epsilon_{x,r})}{(1-\eta_r-\epsilon_{x,r}+\eta_x)(3-2\eta_x-\epsilon_{r,x})},
\end{equation}

\begin{equation}\label{eq:Tr}
T_r=\frac{|\Wset_r|(1-2\eta_r-\epsilon_{x,r}+2\eta_x)+|\Wset_x|(1-\epsilon_{x,r})}{(1-\eta_r-\epsilon_{x,r}+\eta_x)(3-2\eta_r-\epsilon_{x,r})},
\end{equation}

\begin{equation}
x=\arg\max_{n \in \Nset} |\Hset_n|=\arg\min_{n \in \Nset} |\Wset_n|,
\end{equation}

\begin{equation}
r=\arg\max_{n \in (\Nset \setminus x)} \frac{|\Wset_n|}{2-\eta_n-\epsilon_{x,n}},
\end{equation}

{\em Proof:} The proof is provided in Appendix C. \hfill $\blacksquare$

\paragraph{Computational Complexity}
In \ncrn ~the source creates a linear combination of all packets with the complexity of $O(M)$ at each transmission slot. Meanwhile, in the local area, for each device the average number of successful receivers is calculated with the complexity of $O(N-1)$. The complexity of calculating the average number of successful receivers for all devices is $O(N^2)$. Then, the device with the maximum average number of successful receivers is selected as the transmitter with the complexity of $O(N)$ and a linear combination of the packets in its {\em Has} set is created with the complexity of $O(M)$ at each transmission slot. By considering the maximum of $M$ transmission slots, the complexity of \ncrn ~is polynomial with the complexity of $O(M^2+N^2)$. This computational complexity, by also taking additional steps such as dividing a file into smaller sets of $M$ packets, makes \ncrn ~applicable for practical deployment.

\subsubsection{\ncin}
In section \ref{sec:ncin_no_loss}, we described and analyzed \ncin ~for the case that the channel links are lossless. In this section, we consider a more generalized case where the channel links are lossy and present \ncin ~for this generalized case.
\paragraph{Algorithm Description}
\ncin ~algorithm, described in  Section \ref{sec:ncin_algorithm}, groups the packets that are wanted by the devices into sets $\Mset_c$, $\Mset_l$, and $\Mset_d$ when there is no loss in the second stage.  Then, packets from these sets are network coded and transmitted from the cellular and D2D links. Yet, when the links are lossy in the second stage, this approach should be revised for the following reasons.

First, each network coded packet is transmitted successfully when there is no loss. This makes creating fixed $\Mset_c$, $\Mset_l$, and $\Mset_d$ sets possible and reasonable before transmitting the packets in the second stage. However, when the channels are lossy in the second stage, the transmitted network coded packet at each transmission slot may be received successfully by some of the targeted receivers (known as successful receivers) and may be lost by the others. Thus, fixed $\Mset_c$, $\Mset_l$, and $\Mset_d$ sets are not appropriate in this scenario, and they should be updated after each transmission depending on successful packet transmissions.

Also, when the channel links are lossy, different network coded packets even if they are in the same set (such as in $\Mset_l$) have different priorities for transmission. In particular, the packets that can be received by possibly larger number of devices successfully should be prioritized as it would deliver more information in one transmission, which would eventually reduce the packet completion time.

By taking into account these two points, we updated \ncin ~for lossy channels. The new algorithm is provided in Algorithm \ref{al:ncin_loss}. In this algorithm, we first determine the sets $\Mset_c$, $\Mset_l$, and $\Mset_d$ using Algorithm \ref{al:Groups} at each slot. Then two packets are selected from these sets; one to be transmitted from the source and the other one to be transmitted by a mobile device using D2D connections. The idea behind packet selection is that the selected packet can be received by possibly largest number of devices successfully as it would deliver more information in one transmission and thus would eventually reduce the packet completion time. The details of packet selection in Algorithm \ref{al:ncin_loss} are described below.

\begin{algorithm}[h]
\caption{\ncin ~for Lossy Channels}
\label{al:ncin_loss}
\begin{algorithmic}[1]
\STATE Group the packets in the {\em Wants} sets of all devices into the sets $\Mset_c$, $\Mset_l$ and $\Mset_d$ using Algorithm \ref{al:Groups} and keep all network coded packets, $p \in (\Mset_c \cup \Mset_l \cup \Mset_d)$, along with their corresponding vectors, $v_p$, in these sets.

{\bf Packet Selection by the Source:}
\IF {$\Mset_c$ is not empty}
\STATE The first element of $\Mset_c$ is selected to be transmitted from the source.
\ELSIF {$\Mset_l$ is not empty}
\STATE The first element of $\Mset_l$ is selected to be transmitted from the source.
\ELSIF {$\Mset_d$ is not empty}
\STATE The packet with the maximum average number of successful receivers among all packets in $\Mset_d$, which is equal to $\arg\max_{p \in \Mset_d}\sum_{n \mid (v_p[n] \neq NULL)} (1-\eta_n)$, is selected to be transmitted from the source.
\ENDIF

{\bf Packet Selection by the Mobile Devices:}
\STATE Consider packet $p$ with its corresponding vector $v_p$ in $\Mset_l$; any device in $x \in \Nset$ can transmit partial of packet $p$ which is constructed by network coding all packets in the vector $v_p \setminus v_p[x]$. The average number of successful receivers for this packet is equal to $\sum_{n \mid (v_p[n] \neq v_p[x])} (1-\epsilon_{x,n})$. Determine all possible transmitted packets with their corresponding transmitters and calculate their average number of successful receivers. \label{step:ml}
\STATE Consider packet $p$ with its corresponding vector $v_p$ in $\Mset_d$; any device that has all uncoded packets in $v_p$ can transmit $p$. The average number of successful receivers for packet $p \in \Mset_d$ to be transmitted from device $x \in \{i \mid v_p[i]=NULL\}$ is equal to $\sum_{n \mid (v_p[n] \neq NULL)} (1-\epsilon_{x,n})$. Determine all possible transmitted packets with their corresponding transmitters and calculate their average number of successful receivers. \label{step:md}
\STATE From all packets determined in \ref{step:ml} and \ref{step:md}, select the packet with the maximum average number of successful receivers to be transmitted from its corresponding transmitter through D2D links.

\end{algorithmic}
\end{algorithm}

{\bf Packet Selection by the Source (lines 2-7):} The average number of successful receivers for any packet in the set $\Mset_c \cup \Mset_l$ that is transmitted from the source is equal to $\sum_{n \in \Nset} (1-\eta_{n})$, because this packet targets all devices. On the other hand, the average number of successful receivers for any packet $p$ in the set $\Mset_d$ with its corresponding vector $v_p$ is equal to $\sum_{n \mid (v_p[n] \neq NULL)} (1-\eta_{n})$, because this packet is innovative for any device $n \in \Nset$ for which $v_p[n]$ is not $NULL$ and thus the set of targeted receivers is $\Nset_r=\{n \mid (v_p[n] \neq NULL)\}$. Since $\Nset_r$ is a subset of $\Nset$, $\sum_{n \in \Nset} (1-\eta_{n})$, the average number of successful receivers for any packet in $\Mset_c$ or $\Mset_l$, is greater than $\sum_{n \in \Nset_r} (1-\eta_{n})$, the average number of successful receivers for any packet in $\Mset_d$ and thus, a packet from the set $\Mset_c$ or $\Mset_l$ is preferred to be transmitted from the source than a packet from $\Mset_d$. Between the sets $\Mset_c$ and $\Mset_l$, a packet from the set $\Mset_c$ is preferred to be selected (lines 2-3), because the packets in $\Mset_c$ are not available to be transmitted through the other interface. Therefore, the order of transmitting the packets from the source is (i) $\Mset_c$, (ii) $\Mset_l$, and (iii) $\Mset_d$. There is no priority among the packets of the same set for the sets $\Mset_c$ and $\Mset_l$, because all packets have the same average number of successful receivers (lines 4-5). For the set $\Mset_d$, the packet with the maximum average number of successful receivers is preferred to be transmitted (lines 6-7).

{\bf Packet Selection by the Mobile Devices (lines 8-10):} We first consider packet $p$ from set $\Mset_l$ with its vector $v_p$. Each device $x \in \Nset$ can transmit a partial of this packet that is available in its {\em Has} set. $v_p[x]$ is the uncoded packet in the network coded packet $p$ that is wanted by device $x$. Therefore, $v_p[x]$ is the only uncoded packet in $v_p$ that is not available in the {\em Has} set of $x$. With that being said, $x$ can transmit the partial of packet $p$ which is constructed by network coding all packets in the vector $v_p \setminus v_p[x]$ with the targeted receivers $\Nset_r = \{i \mid (v_p[i] \neq v_p[x])\}$. Therefore, the average number of successful receivers for the partial packet transmitted from $x$ is equal to $\sum_{n \mid (v_p[n] \neq v_p[x])} (1-\epsilon_{x,n})$. We determine all possible transmitted packets from the set $\Mset_l$ along with their corresponding transmitters and calculate their average number of successful receivers (line 8).
Then, we consider packet $p$ from set $\Mset_d$ with its vector $v_p$. Each device $x$ for which $v_p[x]=NULL$ can transmit packet $p$, because it has all uncoded packets of $p$ in its {\em Has} set and the set of its targeted receivers is $\Nset_r=\{i \mid (v_p[i] \neq NULL)\}$. With that being said, the average number of successful receivers for packet $p \in \Mset_d$ transmitted from $x$ is equal to $\sum_{n \mid (v_p[n] \neq NULL)} (1-\epsilon_{x,n})$. We determine all possible transmitters for each packet from set $\Mset_d$ and calculate their average number of successful receivers (line 9).
At last, we select the packet with the maximum average number of successful receivers among all packets from the sets $\Mset_l$ and $\Mset_d$ to be transmitted from its corresponding transmitter through D2D links (line 10).

\paragraph{Upper Bound on $T$}
In order to characterize the performance of proposed \ncin, we develop the following upper bound for the packet completion time obtained from \ncin ~for the lossy channels in the second stage.

\begin{theorem} \label{th:upper_bound_lossy}
The average of packet completion time; $T$ when \ncin ~is employed by cooperative mobile devices on a joint cellular and D2D setup when the channel links are lossy is upper bounded by

\begin{equation} \label{eq:b2_instant}
T \leq \left\lceil \max \bigg(T_{s,c},\frac{(T_{l,d}+T_{l,l})(T_{s,c}+T_{s,d}+T_{s,l})}{T_{l,d}+T_{l,l}+T_{s,d}+T_{s,l}}\bigg) \right\rceil,
\end{equation}
\end{theorem}
where $\Mset_c$, $\Mset_l$, and $\Mset_d$ are the sets that are constructed by Algorithm \ref{al:Groups} for the first transmission slot and $T_{s,c}$, $T_{s,l}$, and $T_{s,d}$ are the average packet completion times for the source to transmit the packets in the sets $\Mset_c$, $\Mset_l$, and $\Mset_d$, respectively. $T_{l,l}$ and $T_{l,d}$ are the average packet completion times for the cooperating devices in the local area to transmit the packets in the sets $\Mset_l$ and $\Mset_d$, as calculated in the following:

\begin{equation} \label{eq:T_sc}
T_{s,c}=\frac{|\Mset_c|}{1-\max\limits_{n \in \Nset} \eta_{n}},
\end{equation}

\begin{equation} \label{eq:T_sl}
T_{s,l}=\frac{|\Mset_l|}{1-\max\limits_{n \in \Nset} \eta_{n}},
\end{equation}

\begin{equation} \label{eq:T_sd}
 T_{s,d}= \sum_{p \in \Mset_d} \frac{1}{1-\max\limits_{n \mid (v_p[n] \neq NULL)} \eta_{n}},
\end{equation}

\noindent where $v_p$ is the vector associated with the network coded packet $p$.

\begin{equation} \label{eq:T_ll}
T_{l,l}= \sum_{p \in \Mset_l} \bigg(\frac{1}{1-\max\limits_{n \mid (v_p[n] \neq v_p[x])} \epsilon_{x,n}} + \frac{1}{1-\max\limits_{n \mid (v_p[n]=v_p[x])} \epsilon_{x',n}}\bigg),
\end{equation}

\noindent where $x$ is the device to be selected to transmit the partial of packet $p \in \Mset_l$. According to Algorithm \ref{al:ncin_loss}, $x$ is selected such that the average number of successful receivers is maximized; $x=\arg\max_{i \in \Nset} \sum_{n \mid (v_p[n] \neq v_p[i])} (1-\epsilon_{i,n})$. $x'$ is the device to be selected to transmit the residual of packet $p$, which includes the uncoded packet $v_p[x]$. $x'$ is selected such that the average number of successful receivers is maximized; $x'=\arg\max_{i \mid (v_p[i] \neq v_p[x])} \sum_{n \mid (v_p[n] = v_p[x])} (1-\epsilon_{i,n})$.

\begin{equation} \label{eq:T_ld}
T_{l,d}= \sum\limits_{p \in \Mset_d} \frac{1}{1-\max\limits_{n \mid (v_p[n] \neq NULL)} \epsilon_{x,n}},
\end{equation}

\noindent where $x$ is the device to be selected to transmit packet $p \in \Mset_d$. According to Algorithm \ref{al:ncin_loss}, $x$ is selected such that the average number of successful receivers is maximized; $x=\arg\max_{i \mid (v_p[i] = NULL)} \sum_{n \mid (v_p[n] \neq NULL)} (1-\epsilon_{i,n})$.

{\em Proof:} The proof is provided in Appendix D. \hfill $\blacksquare$

\paragraph{Computational Complexity}
In \ncin ~Algorithm \ref{al:Groups} is run with the complexity of $O(M^2+N)$, at each transmission slot. Then, the source calculates the average number of successful receivers (if it is required to transmit a packet from the set $\Mset_d$) with the complexity of $O(MN)$, selects the packet with maximum average number of successful receivers with the complexity of $O(M)$, and creates a network coded packet with the complexity of $O(M)$ at each transmission slot. Meanwhile, in the local area, the average number of successful receivers is calculated for all devices as the transmitter and all packets with the complexity of $O(MN^2)$, the packet with the maximum average number of successful receivers with its corresponding transmitter is selected with the complexity of $O(MN)$, and a network coded packet with the complexity of $O(M)$ is created, at each transmission slot. Therefore, the complexity of \ncin ~at each transmission slot is $O(M^2+N^2)$. By considering the maximum of $M$ transmission slots, the complexity of \ncin ~is polynomial with the complexity of $O(M^3+N^2)$. This computational complexity, by also taking additional steps such as dividing a file into smaller sets of $M$ packets, makes both \ncin ~applicable for practical deployment.

\vspace{-5pt}
\section{\label{sec:lower} Lower Bound on $T$}
\vspace{-5pt}
In this section, we develop a lower bound on the packet completion time when any network coding algorithm is employed by cooperative mobile devices on a joint cellular and D2D setup.\footnote{Note that the provided lower bound is not guaranteed to be achievable. However, it is guaranteed that there is no other scheme to perform better than the lower bound and thus it is a good metric to evaluate the performance of proposed NCMI methods.} The effectiveness of a network coding algorithm is evaluated by comparing the packet completion time obtained from the algorithm with the lower bound; the closer the packet completion time obtained from a network coding algorithm is to the lower bound, the more effective is the algorithm.

\begin{theorem} \label{th:lowerbound_loss}
The packet completion time when network coding is employed by cooperative mobile devices on a joint cellular and D2D setup is lower bounded by:
\begin{equation} \label{eq:lower_loss}
T \geq \left \lceil \max \bigg(\frac{|\Mset_c|}{1-\prod_{n \in \Nset}\eta_n},\max_{n \in \Nset}\big(\min_{x \in (\Nset \setminus n)}\frac{|\Wset_n|}{2-\eta_n-\epsilon_{x,n}}\big)\bigg) \right \rceil.
\end{equation}
\end{theorem}

{\em Proof:} The proof is provided in Appendix E. \hfill $\blacksquare$

\begin{corollary} \label{th:lowerbound}
The packet completion time when network coding is employed by cooperative mobile devices on a joint cellular and D2D setup when the channel links are lossless in the second stage, is lower bounded by:
\begin{equation} \label{eq:lower}
T \geq \left \lceil \max\bigg(|\Mset_c|,\frac{1}{2}{\max_{n \in \Nset} |\Wset_n|}\bigg) \right \rceil.
\end{equation}
\end{corollary}
{\em Proof:} By substituting $\eta_n=\epsilon_{k,l}=0, \forall n,k,l \in \Nset$ in (\ref{eq:lower_loss}), the lower bound provided in (\ref{eq:lower}) is obtained. \hfill $\blacksquare$

As shown in the simulation results, our proposed \ncrn ~and \ncin ~methods perform closer to the lower bound as compared to the baselines.

\vspace{-5pt}
\section{\label{sec:simulation}Simulation Results}
\vspace{-5pt}
We considered a topology shown in Fig.~\ref{fig:intro_example}(b) for different number of devices, packets, and loss probabilities. Then we implemented our proposed methods, \ncrn ~and \ncin ~for this topology and compared their performances with the lower bound and their upper bounds as well as the baselines of, (i) NoNC, which stands for No Network Coding scheme and (ii) NCSI, which stands for Network Coding for Single Interface systems. Each simulated point, is the average of results over 500 iterations. In our simulation results, bounds are plotted using dashed lines, while the real simulation results are plotted using the solid lines. We first consider the case where the channels are lossless in the second stage. Then, we present the simulation results for lossy channels in the second stage. Finally, we investigate the effect of subpacketization on both lossless and lossy NCMI.

\vspace{-5pt}
\subsection{Lossless Channels in Stage Two}
\begin{figure}[t!]
\centering
\subfigure[Packet completion time vs. number of packets]{ \scalebox{.3}{\includegraphics{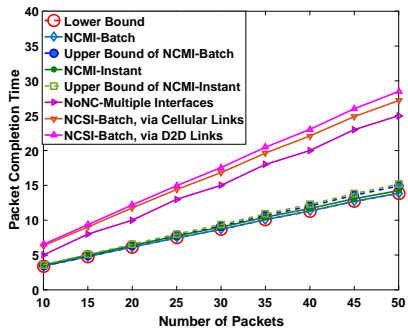}} }
\subfigure[Zoomed version of (a)]{ \scalebox{.3}{\includegraphics{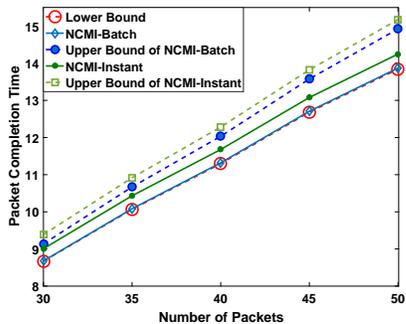}} }
\caption{(a) Packet completion time vs. number of packets for \ncin ~and \ncrn ~as compared to the lower bound and their upper bounds as well as baselines, when the channels are lossless in stage two. (b) Zoomed version of (a).}
\label{fig:sim_pack}
\end{figure}

\begin{figure}[t!]
\centering
{ \scalebox{.3}{\includegraphics{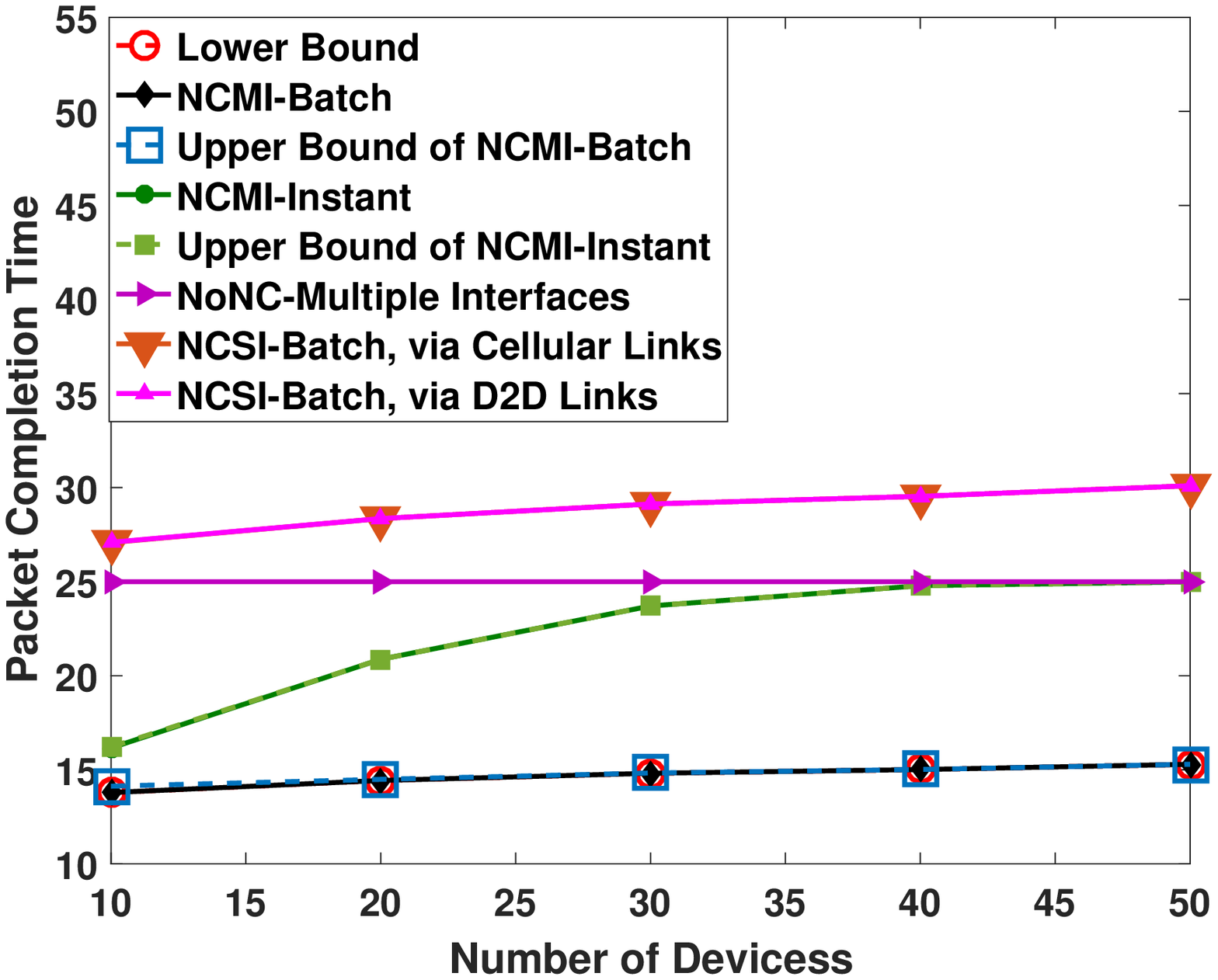}} }
\caption{Packet completion time vs. number of devices for \ncin ~and \ncrn ~as compared to the lower bound and their upper bounds as well as baselines, when the channels are lossless in stage two.}
\label{fig:sim_user}
\end{figure}

\begin{figure}[t!]
\centering
\scalebox{0.3}{\includegraphics{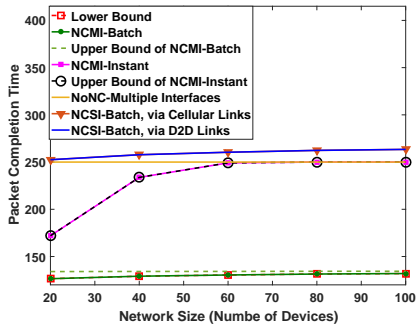}}
\caption{Scalability of \ncin ~and \ncrn ~with the network size as compared to the lower bound and their upper bounds as well as baselines, when the channels are lossless in stage two.}
\label{fig:CT_vs_NS}
\end{figure}

In this section, we consider a setup, in which each device misses the packets transmitted from the source in stage one with a specific loss probability. The loss probability for each device in stage one, is selected uniformly from $[0.3, 0.5]$ for Figs. \ref{fig:sim_pack}, \ref{fig:sim_user}, and \ref{fig:CT_vs_NS}. Note that the number of lost packets at the start of stage two is equal to $M=|\bigcup_{n \in \mathcal{N}}W_n|$. All packet transmissions in stage two are assumed to be lossless. We implemented our proposed schemes in this setup, \ncrn ~and \ncin, and compared their packet completion time performances with: (i) the {\em Lower Bound}, in Eq. \ref{eq:lower}, (ii) the {\em Upper Bounds} provided in Eq. \ref{eq:b2_random} for \ncrn ~and Eq. \ref{eq:better_upper} for \ncin, (iii) {\em NoNC-Multiple Interfaces}, which is a no network coding scheme, but using cellular and D2D links jointly,
(iv) {\em NCSI-Batch, via Cellular Links}, which uses batch-based network coding via the single interface of cellular links,
(v) {\em NCSI-Batch, via D2D Links}, which uses batch-based network coding via the single interface of D2D links. Note that packets in $\Mset_c$ are requested from the source device via the cellular links in {\em NCSI-Batch, via D2D Links} scheme.

\begin{figure}[t!]
\centering
{ \scalebox{.3}{\includegraphics{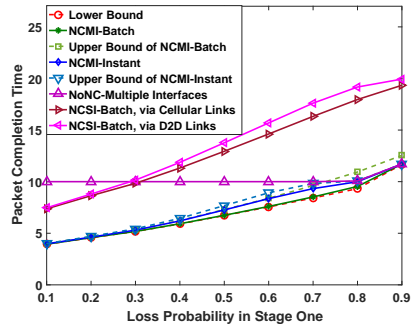}} }
\caption{Packet completion time vs. loss probabilities of cellular links in stage one for \ncin ~and \ncrn ~as compared to the lower bound and their upper bounds as well as baselines, when the channels are lossless in stage two.}
\label{fig:sim_loss}
\end{figure}

{\bf Packet completion time vs. number of packets:} Fig. \ref{fig:sim_pack} shows the packet completion time for different number of packets and $N=5$ devices. As seen, \ncin ~and \ncrn ~improve the packet completion time significantly as compared to the single-interface systems and No-NC. This shows the effectiveness of (i) using multiple interfaces as compared to the single-interface systems, and (ii) employing network coding. \ncrn ~and \ncin ~are slightly better than their upper bounds for larger number of lost packets, because the upper bounds give the worst case performance guarantee for \ncrn ~and \ncin. Yet, as seen in the zoomed version of the figure (\ie Fig. \ref{fig:sim_pack}(b)), the upper and lower bounds are very close to the actual performance of \ncrn ~and \ncin. This shows the tightness of our upper and lower bound analysis.

Also, as seen in Fig. \ref{fig:sim_pack}(a), single interface cellular system has less packet completion time than single interface D2D system. The reason is that any network coded packet can be created in the source and transmitted through the cellular link, while in the local area, only the network coded packets that can be created in one of the devices can be selected to be transmitted. Therefore, the network coded packet transmitted through the cellular links may target larger number of devices than the packets that are transmitted through D2D links and thus results in less packet completion time.

{\bf Packet completion time vs. number of devices:} Fig. \ref{fig:sim_user} shows the packet completion time for different number of devices and $M=50$ packets. As shown in the figure, \ncrn ~performs very close to the lower bound and upper bounds, and better than the single-interface systems and No-NC. \ncin ~performs better than single-interface systems and No-NC and worse than \ncrn. The performance of \ncin ~gets closer to No-NC method for the larger number of devices. The reason is that the network coding opportunities of NCMI-Instant decrease in this lossless scenario when the number of devices increases, so NCMI-Instant gets closer to NoNC.

{\bf Packet completion time vs. loss probability:} Fig. \ref{fig:sim_loss} presents the packet completion time for different loss probabilities in stage one for $M=20$ packets and $N=5$ devices. In this setup, the loss probabilities of cellular links are the same for all mobile devices. As seen, the performances of {\tt NCMI} schemes and the lower bound get closer to the no network coding scheme by increasing the channel losses of cellular links. In other words, when the channel losses are large, the no network coding scheme (with lower complexity than network coding schemes) has the performance close to the optimal scheme, which is defined by the lower bound.

\begin{figure*}[!h]
\centering
\subfigure[
Packet completion time vs. number of packets]{ \scalebox{.2}{\includegraphics{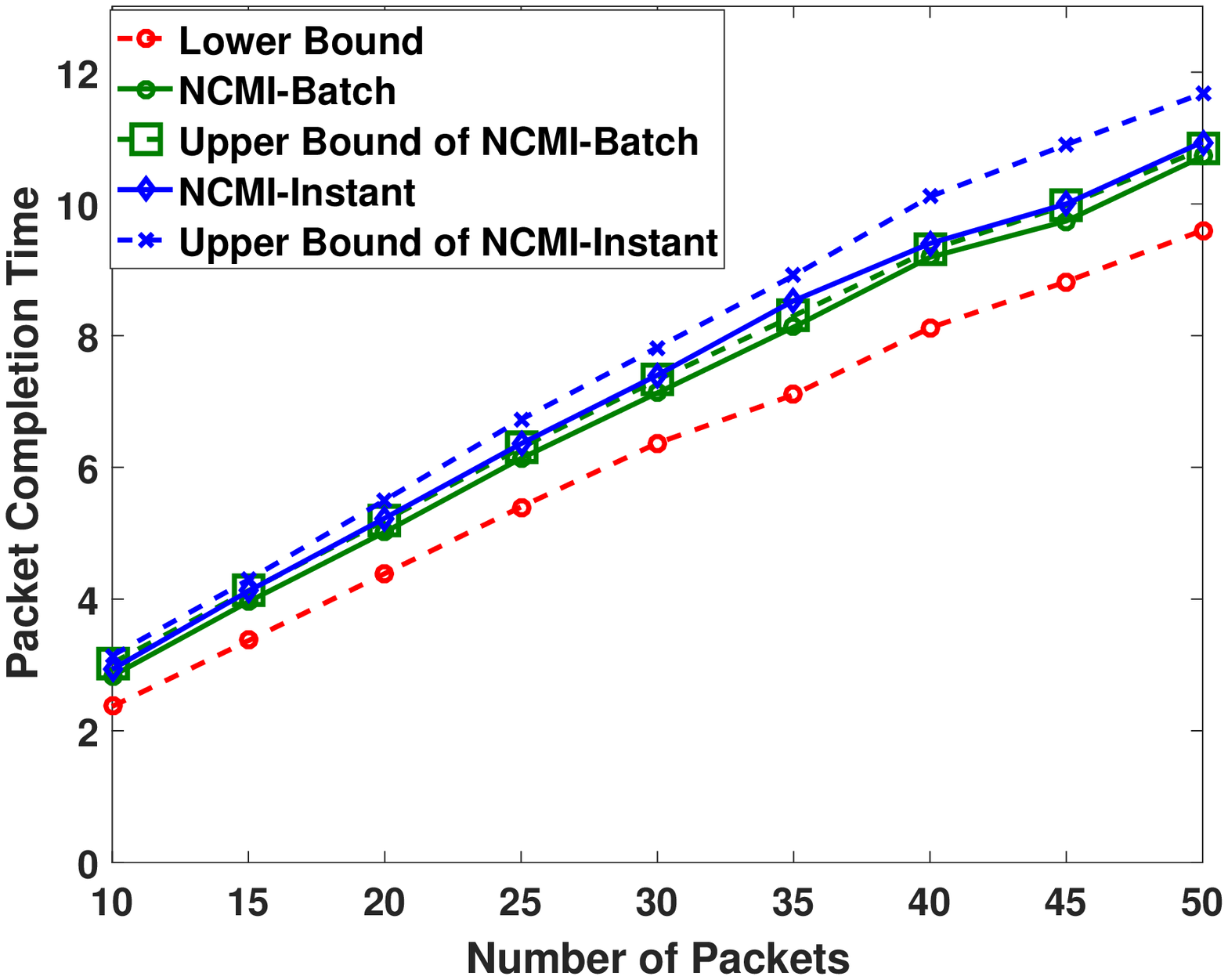}} }
\subfigure[
Packet completion time vs. number of packets]{ \scalebox{.2}{\includegraphics{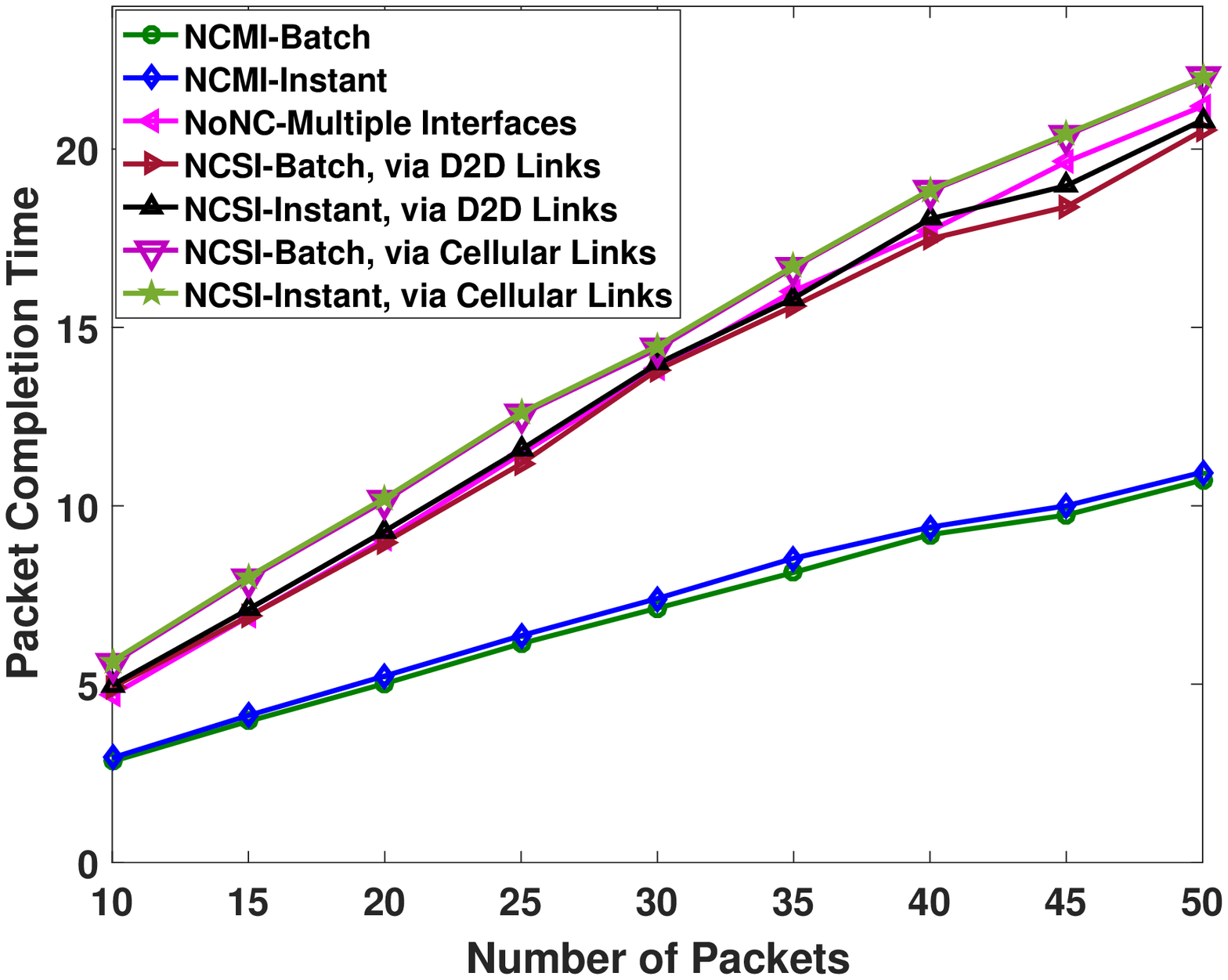}} }
\subfigure[
Packet completion time vs. number of devices]{ \scalebox{.2}{\includegraphics{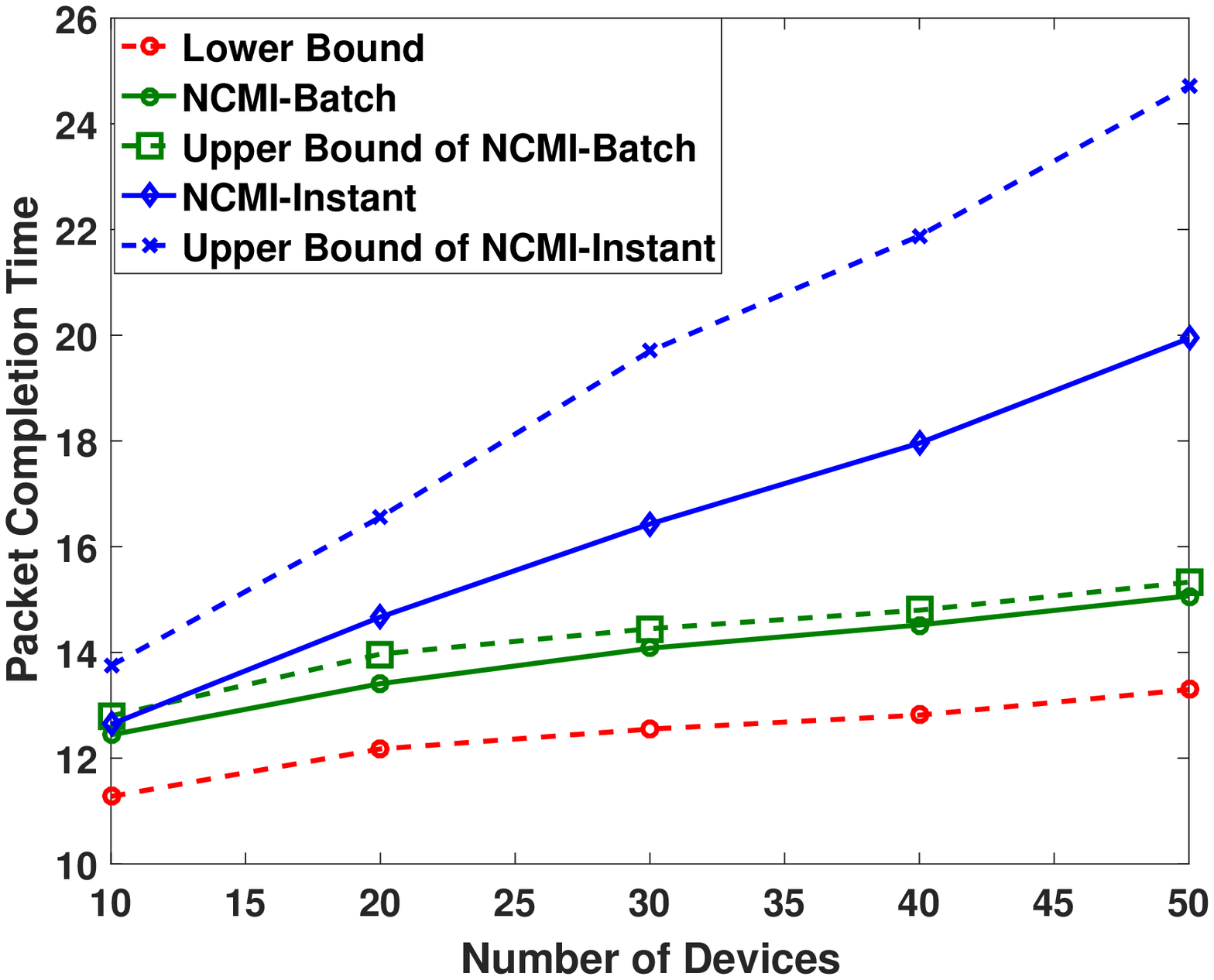}} }
\subfigure[
Packet completion time vs. number of devices]{ \scalebox{.2}{\includegraphics{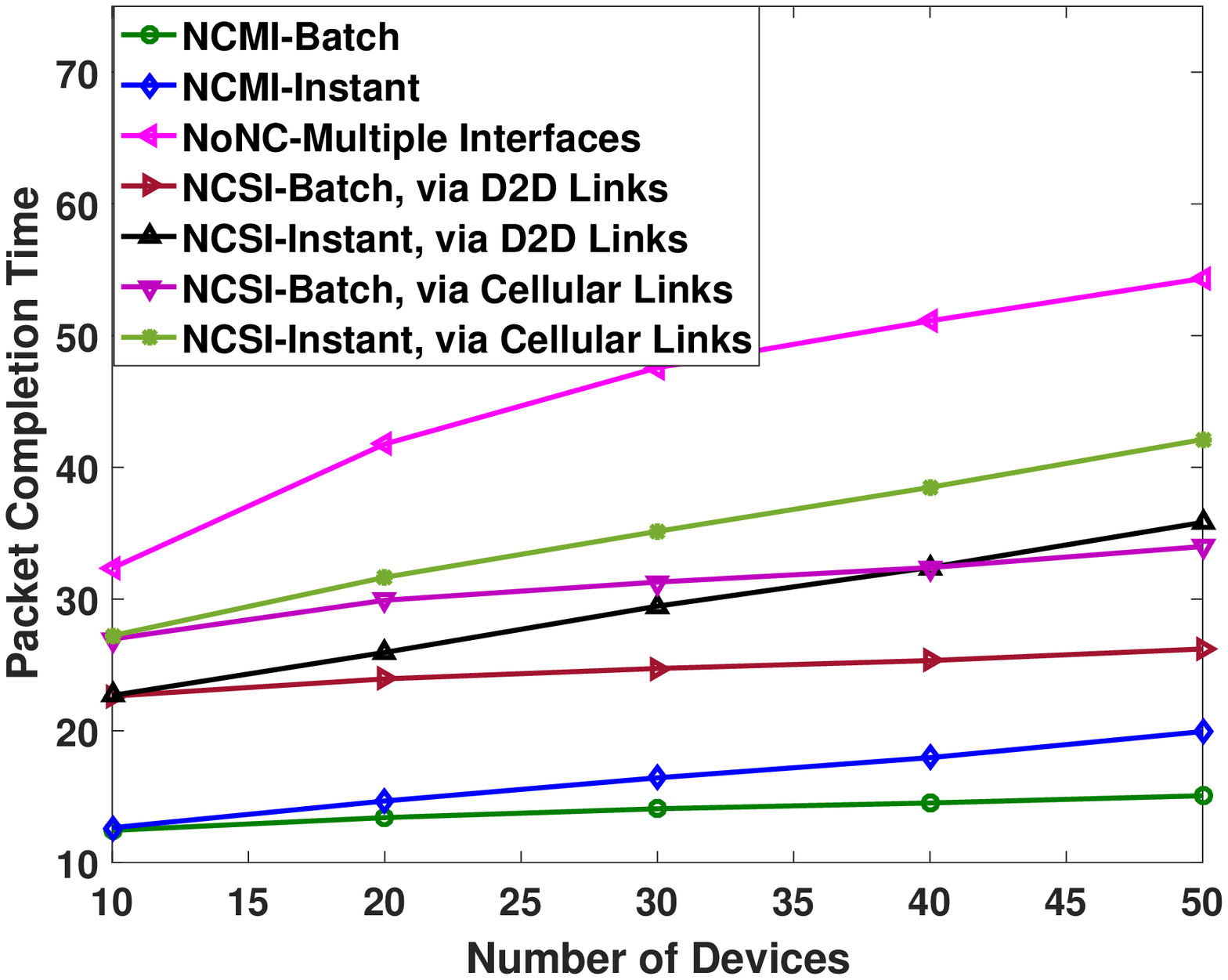}} }
\caption{
The packet completion time performance of \ncin ~and \ncrn ~as compared to the lower bound and the baselines, when the channels are lossy in stage two.}
\label{fig:sumulations_loss}
\vspace{-10pt}
\end{figure*}

{\bf Scalability with the Network Size:} Fig. \ref{fig:CT_vs_NS} shows the packet completion time for different network sizes (number of packets) with $M=500$ packets. As seen, \ncrn ~scales well with increasing the network size, but \ncin ~performs close to No-NC scheme for larger networks and fixed number of packets. In order for \ncin ~to perform close to its best performance, the number of packets should also be increased by increasing the size of the network. The reason for this observation is that network coding opportunities of \ncin ~decrease with increasing number of devices in lossless scenario (lossless in stage two). This holds for a general setup as explained next.

Let us assume that we would like to create an IDNC packet $p_m + p_k$. The condition to create this IDNC packet is that all devices that want $p_m$ should have packet $p_k$, and all devices that want $p_k$ should have $p_m$. The probability of this condition is $\prod_{n \in \Nset_r}(1-\eta_n)$, where $\Nset_r$ is the set of devices that want packet $p_m$ or $p_k$ and $\eta_n$ is the loss probability at stage one. As seen, the probability of creating an IDNC packet, \ie $\prod_{n \in \Nset_r}(1-\eta_n)$ decreases with increasing $|\Nset_r|$.

\subsection{Lossy Channels in Stage Two}
In this section, we consider a setup in which the packet transmissions in both stages are lossy. In stage one, all packets are transmitted from the source; each transmitted packet is lost with the loss probability of $\eta_n$ at each device $n$. In stage two, two packets are broadcast simultaneously at each transmission slot; one packet is broadcast from the source and is lost with the loss probability of $\epsilon_n$ at each device $n$ and the other packet is broadcast from the transmitter device $t$ (to all other devices) and is lost with the loss probability of $\epsilon_{t,n}$ at each device $n$. The channel loss probabilities $\eta_n, n \in \Nset$ and $\epsilon_{t,n}, t,n \in \Nset$ (in both stage one and stage two) are selected uniformly from $[0.15, 0.35]$ for Figs. \ref{fig:sumulations_loss} and \ref{fig:Subfile}.

We implemented our proposed schemes, \ncrn ~and \ncin ~in this setup and compared their packet completion time performances with the following: (i) {\em Lower Bound}, which is derived in (\ref{eq:lower_loss}). (ii) {\em Upper Bounds}, provided in (\ref{eq:b2_random_l}) for \ncrn ~and (\ref{eq:b2_instant}) for \ncin. (iii) {\em NoNC-Multiple Interfaces}, which is a no network coding scheme, but using cellular and D2D links jointly. At each transmission slot, each interface selects a missing packet with the maximum average number of receivers to be broadcast to all devices. (iv) {\em NCSI-Batch, via D2D Links}, which uses instantly decodable network coding via the single interface of D2D links. At each transmission slot, the device with the maximum size of {\em Has} set is selected as the transmitter and broadcast a random combination of all packets in its {\em Has} set to all other devices. (v) {\em NCSI-Instant, via D2D Links}, which uses instantly decodable network coding via the single interface of D2D links. At each transmission slot, packets are grouped to $\Mset_c$, $\Mset_l$, and $\Mset_d$ sets according to Algorithm \ref{al:Groups}. Then, if $\Mset_c$ is not empty, the packets in this set are requested to be transmitted from the source until they are received successfully by at least one of the devices. If $\Mset_c$ is empty, the order of transmitting the packets is from the sets $\Mset_d$ and $\Mset_l$; at each transmission slot, the packet with the maximum average number of successful receiver is selected among all packets of the same set.
(vi) {\em NCSI-Batch, via Cellular Links}, which uses batch-based network coding via the single interface of cellular links. At each transmission slot, a random combination of all packets is broadcast to all devices from the source.
(vii) {\em NCSI-Instant, via Cellular Links}, which uses instantly decodable network coding via the single interface of cellular links. At each transmission slot, the union of the missing packets in all devices are grouped into $\Mset_c$, $\Mset_l$, and $\Mset_d$ sets according to Algorithm \ref{al:Groups}. The order of transmitting the packets is from the sets $\Mset_c$, $\Mset_l$ and $\Mset_d$.

{\bf Packet completion time vs. number of packets/number of devices:} Figs. \ref{fig:sumulations_loss}(a) and (b) show the packet completion time for different number of packets and $N=5$ devices and Figs. \ref{fig:sumulations_loss}(c) and (d) show the packet completion time for different number of devices and $M=50$ packets. As seen, \ncrn ~and \ncin ~improve the packet completion time significantly as compared to the single-interface systems and No-NC. ote that \ncin ~significantly improves over NoNC when the number of devices increases. This is different than what we observed in Fig. \ref{fig:sim_user} (where \ncin ~gets closer to NoNC with increasing number of devices). The reason is that network coding opportunities of \ncin ~still exist in this scenario due to losses in stage two. Also, network coding schemes using the single interface of D2D links outperform the network coding schemes using single interface of cellular links, despite the fact that more various network coded packet can be transmitted from the cellular link than the D2D link. The reason is that the packets to satisfy each device $n$ using D2D links can be transmitted from a variety of channel links with the loss probabilities of $\epsilon_{t,n}, t \in (\Nset \setminus n)$ from which the most reliable link is selected. While the packets to satisfy each device $n$ using the cellular links can only be transmitted from the cellular link with the fixed loss probability of $\eta_n$ and thus if $\eta_n$ is high, there is no other more reliable link to be selected. Finally, the performance of \ncin ~and \ncrn ~are close to the lower bound and their upper bounds.

\subsection{Effect of Subpacketization on Lossless and Lossy NCMI}
\begin{figure}[t!]
\centering
\subfigure[
Packet completion time vs. subfile size for lossless channels]{ \scalebox{.35}{\includegraphics{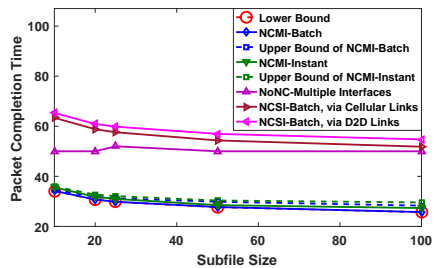}} }
\subfigure[
Packet completion time vs. subfile size for lossy channels]{ \scalebox{.35}{\includegraphics{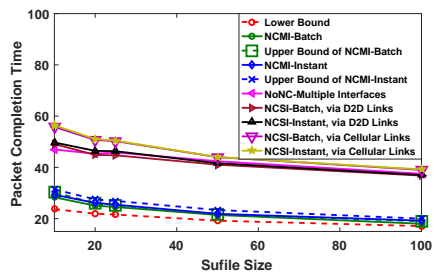}} }
\caption{
The impact of subpacketization on the packet completion time performance.
}
\label{fig:Subfile}
\end{figure}

We investigated the impact of subpacketization for a file with the size of $M=100$ packets and $N=5$ cooperating devices, and showed the results in Fig. \ref{fig:Subfile}. In this figure, a file with fixed size of $M=100$ packets is transmitted for all simulated points, and the packet completion time is measured for this file. In particular, Fig. \ref{fig:Subfile} presents the packet completion time of the file with $M=100$ packets when this file is divided into different sized subfiles. For example, in Fig. \ref{fig:Subfile} (b), the packet completion time of NCMI is close to 30 when subfile size is $10$. This means that (i) the file of $M=100$ packets is divided into 10 subfiles, each with 10 packets, (ii) each subfile is coded using NCMI, and (iii) the total delay of 10 subfiles (\ie the whole file with $M=100$ packets) is reported. The same argument holds for all of the simulated points in Fig. \ref{fig:Subfile}.

As seen in the figure, the packet completion time decreases with increasing the size of subfile. This means that it would be more efficient to apply proposed methods on the whole file instead of dividing it into subfiles and apply the proposed methods on each subfile. As seen, our NCMI algorithms still significantly improve the packet completion time as compared to the baselines.

\section{\label{sec:related} Related Work}
{\em Network Coding for Single-Interface Systems:} The performance of network coding has been evaluated for single-interface systems in literature. 
The problem of minimizing the number of broadcast transmissions required to satisfy all devices is considered in \cite{salim_broadcast}, and the bounds for packet completion time are developed. A deterministic linear network coding algorithm that minimizes the number of broadcast transmissions is considered in  \cite{parastoo_broadcast}.  Minimization of the completion delay while broadcasting instantly decodable network coding packets has been considered in \cite{sorourICC}. The problem of recovering the missing content using cooperative data exchange utilizing local area connections is considered in \cite{RouayhebITW10} and \cite{RouayhebISIT10}, and the lower and upper bounds on the minimum number of transmissions are developed. Deterministic algorithms for the cooperative data exchange problem with polynomial time are designed in \cite{RouayhebISIT11} and \cite{SprintsonQShine10}. In our previous work, \cite{CA_IDNC}, we considered data exchange problem for multimedia applications and proposed content-aware network coding schemes that improves content quality and reduces the packet completion time using D2D links only. As compared to \cite{CA_IDNC}, we consider cooperative mobile devices in the joint cellular and D2D setup, and develop a network coding scheme for this setup in this paper.

{\em Multiple-Interface Systems:}
The performance of WiFi-only, cellular-only, and multiple-interface (WiFi plus cellular) systems are studied and compared in \cite{MI1}. A flexible software architecture is developed in \cite{MI2} by adaptively selecting available interfaces at mobile devices with the goal of improving Quality of Experience (QoE) while minimizing the energy overhead. The heterogeneity of cellular and Wi-Fi interfaces are effectively utilized to deliver data to mobile devices in \cite{MI3}. Cooperative content delivery to mobile devices is developed in \cite{MI4} by taking into account the link quality of both cellular and WiFi interfaces, where a device with the best link quality is preferred in the cooperative system. The scenario of delivering same content to a group of users by using unicast cellular and D2D links is considered in \cite{MI5}.
A comprehensive survey on exploiting multiple available wireless interfaces to deliver content to cooperative mobile devices is provided in \cite{MI7}. This line of work demonstrates the practicality of simultaneous operation of multiple interfaces including cellular and short-range D2D links. As compared to this work, we consider how efficient network coding algorithms can be developed with provable performance guarantees for cooperative mobile devices in the joint broadcast cellular and broadcast D2D setup.

{\em Network Coding for Multiple-Interface Systems:} Network coding has been employed in the previous work for devices with multiple interfaces. Wireless video broadcasting with P2P error recovery is proposed by Li and Chan \cite{micro18}. An efficient scheduling approach with network coding for wireless local repair is introduced by Saleh et al. \cite{micro19}. Another body of work \cite{micro20,micro21,micro22} proposes systems where there are a base station broadcasting packets and a group of smartphone users helping each other to correct errors. Compared to prior work \cite{micro18,micro19,micro20,micro21,micro22}, where each phone downloads all the data, and D2D links are used for error recovery, our scheme jointly utilizes cellular and D2D links and analyzes the performance of network coding in such a setup.

Simultaneous operation of multiple interfaces and employing network coding for this setup has also been considered in the previous work; \cite{microcast,microcast_allerton,AliParan11}.
As compared to \cite{microcast} and \cite{microcast_allerton}, we consider broadcast cellular links simultaneously with D2D links for error recovery purposes, while \cite{microcast} and \cite{microcast_allerton} use unicast cellular links simultaneously with D2D links for throughput improvement purposes. Multimedia streaming to a single user is considered in [25], where multiple interfaces are used at the single user. As compared to this work, we use cooperation among mobile devices that benefit D2D links in conjunction with cellular links. Also, in this paper, we consider how efficient network coding algorithms can be developed with provable performance guarantees for cooperative mobile devices in the joint cellular and D2D setup, instead of using existing network coding algorithms. In \cite{milcom}, the conference version of this work, we developed network coding schemes for cooperative mobile devices in the joint cellular and D2D setup, where we assumed the data transmissions are lossless. As compared to \cite{milcom}, in this paper, we consider lossy channels and propose network coding schemes by taking into account the probabilities of channel losses as well.

\section{\label{sec:conclusion}Conclusion}
In this paper, we considered a scenario where a group of mobile devices is interested in the same content, but each device has a partial content due to packet losses over links. In this setup, mobile devices cooperate and exploit the cellular and D2D links jointly to recover the missing content. We developed network coding schemes; \ncrn ~and \ncin ~for this setup, and analyzed their packet completion time. Simulation results confirm that \ncrn ~and \ncin ~significantly reduce the packet completion time.

\clearpage
\newpage

\newpage
\section*{\label{appendix_upper_batch}Appendix A: Proof of Theorem~\ref{th:easy_upper_bound}}
In \ncrn, at each transmission slot, a linear combination of all missing packets, that is innovative for all mobile devices, is broadcast from the source to all devices. Meanwhile, in the local area, the device with the largest {\em Has} set is selected as the transmitter; the transmitter broadcasts a linear combination of the packets in its {\em Has} set, that is innovative for all other devices. Therefore, the size of {\em Has} set for all devices except for the device with the maximum size of {\em Has} set (the transmitter) is increased by two and the size of {\em Has} set for the transmitter is increased by one.
In the first transmission slot, $x = \arg \max_{n \in \Nset} |\Hset_n|$ is selected as the transmitter so the size of $\Hset_n, n \in (\Nset \setminus x)$ is increased by two and the the size of $\Hset_{x}$ is increased by one.
For the next transmission slots, $x$ remains the transmitter until the size of {\em Has} set for one of the other devices is equal to the size of the {\em Has} set for $x$. It takes at most $|\Hset_x|-|\Hset_n|$ transmission slots for device $n$ to have the same size of {\em Has} set as device $x$ and thus to be selected as the transmitter. Therefore, by considering $r = \arg\min_{n \in (\Nset \setminus x)} |\Hset_n| = \arg\max_{n \in (\Nset \setminus x)} |\Wset_n|$, it takes at most $|\Hset_t|-|\Hset_r|$ transmission slots for the transmitter to be reselected. On the other hand, it takes at most $M-|\Hset_x|$ transmission slots that the size of $\Hset_{x}$ becomes equal to $M$. We consider two cases:

\begin{enumerate}

\item $(M- |\Hset_x|) \leq |\Hset_x|-|\Hset_r|$:

After at most $M-|\Hset_x|$ transmission slots, the size of $\Hset_x$ becomes equal to $M$ and the size of $\Hset_r$ is still less than the size of $\Hset_x$. Therefore, in the rest of the transmission slots, $x$ remains as the transmitter and thus, the packet completion time is upper bounded by the required number of transmission slots to satisfy device $r$, which is equal to $\frac{1}{2} \max_{n \in \Nset}|\Wset_n|$.

\item $(M- |\Hset_{x}|) \ge |\Hset_{x}|-|\Hset_r|$:

After at most $|\Hset_{x}|-|\Hset_r|$ transmission slots, the size of $\Hset_{x}$ becomes equal to $\Hset_r$. Therefore, in the rest of the transmission slots, $r$ and $x$ are selected as the transmitter, alternatively; \ie in one of the transmission slots $x$ is selected as the transmitter and in the consecutive transmission slot, $r$ is selected as the transmitter and thus in every two consecutive transmission slots, the size of $\Hset_r$ is increased by three. The packet completion time is upper bounded by the required number of transmission slots to satisfy device $r$, which is equal to $\frac{1}{3} (|\Wset_r|+|\Wset_{x}|)$.

\end{enumerate}

In addition, the number of transmission slots cannot be less than $|\Mset_c|$. By considering this fact and the results from cases (i) and (ii), the upper bound in Theorem~\ref{th:easy_upper_bound} is obtained. This concludes the proof. 

\section*{\label{appendix_upper_instant}Appendix B: Proof of Theorem~\ref{theorem:better_upper}}
We consider three conditions based on the relative sizes of the sets $\Mset_c$, $\Mset_l$ and $\Mset_d$ and then calculate the maximum packet completion time obtained from each of the conditions.

\begin{enumerate}
\item{$|\Mset_c| \geq (|\Mset_d|+2|\Mset_l|)$}\\
Under this condition, the base station starts transmitting the packets in $\Mset_c$; meanwhile, in the local area a network coded packet in $\Mset_d$ and $\Mset_l$ with the order of (i) $\Mset_d$ and (ii) $\Mset_l$ is selected to be transmitted from one of the mobile devices. After $|\Mset_d|+2|\Mset_l|$ transmission slots, all the packets in $\Mset_d$ and $\Mset_l$ are transmitted by the cooperating devices and $|\Mset_c|-(|\Mset_d|+2|\Mset_l|)$ packets are left from $\Mset_c$; it takes $|\Mset_c|-(|\Mset_d|+2|\Mset_l|)$ transmission slots for the base station to transmit these remaining packets. By summing the required number of transmission slots, the packet completion time under condition (1) is equal to:

\begin{equation}
T_{(1)}=|\Mset_c|
\end{equation}

\begin{example}
Let us consider three mobile devices with the {\em Wants} sets of $\Wset_A=\{p_1,p_2,p_3,p_4,p_5\}, \Wset_B=\{p_1,p_2,p_3,p_4,p_6,p_7\}, \Wset_C=\{p_1,p_2,p_3,p_4,p_6,p_8\}$. By using Algorithm \ref{al:Groups}, $\Mset_c=\{p_1,p_2,p_3,p_4\}, \Mset_l=\{p_5+p_6\}, \Mset_d=\{p_7+p_8\}$. For this example, condition (1) is met; $|\Mset_c|=4>3=(|\Mset_d|+2|\Mset_l|)$. Accordingly, $4$ transmission slots are required; in the first transmission slot, $p_1$ is transmitted from the base station and at the same time $p_7+p_8$ is transmitted from device $A$. In the second transmission slot, $p_2$ is transmitted from the base station and $p_6$ is transmitted from device $A$. In the third transmission slot, $p_3$ is transmitted from the base station and $p_5$ is transmitted from device $B$ (or device $C$). In the forth transmission slot, $p_4$ is transmitted from the the base station.
\hfill $\Box$
\end{example}

\item{$(|\Mset_d|+2|\Mset_l|) \geq |\Mset_c| \geq (|\Mset_d|-|\Mset_l|)$}\\
For this condition, we consider two cases of (i) $|\Mset_c| \leq |\Mset_d|$ and (ii) $|\Mset_c| \geq |\Mset_d|$.

In case (i), the base station starts transmitting the packets in $\Mset_c$; meanwhile in the local area the packets in $\Mset_d$ are transmitted from $x$ (the device that has all packets in $\Mset_d$). Since $|\Mset_c| \leq |\Mset_d|$, after $|\Mset_c|$ transmission slots, all the packets in $\Mset_c$ have been transmitted by the base station and $|\Mset_d|-|\Mset_c|$ packets are left from $\Mset_d$. According to condition (2), $|\Mset_c|$ is greater than $(|\Mset_d|-|\Mset_l|)$ and thus $(|\Mset_d|-|\Mset_c|)$ is smaller than $|\Mset_l|$. Therefore, in the next $|\Mset_d|-|\Mset_c|$ transmission slots, the remaining packets in $\Mset_d$ is transmitted by the cooperative devices in the local area and the base station transmits the network coded packets from $\Mset_l$. At last, $|\Mset_l|-(|\Mset_d|-|\Mset_c|)$ packets are left from $\Mset_l$; it takes $2/3(|\Mset_l|-(|\Mset_d|-|\Mset_c|))$ transmission slots by the source and the cooperating devices, jointly to transmit these remaining packets. By summing the required number of transmission slots, the packet completion time for case (i) is equal to $\frac{1}{3}(2|\Mset_l|+2|\Mset_c|+|\Mset_d|)$.

In case (ii), the base station starts transmitting the packets in $\Mset_c$; meanwhile in the local area the packets in $\Mset_d$ are transmitted from $x$ (the device that has all packets in $\Mset_d$). Since $|\Mset_c| \geq |\Mset_d|$, after $|\Mset_d|$ transmission slots, all the packets in $\Mset_d$ have been transmitted by the cooperating devices in the local area and $|\Mset_c|-|\Mset_d|$ packets are left from $\Mset_c$. According to condition (2), $|\Mset_c|$ is smaller than $(|\Mset_d|+2|\Mset_l|)$ and thus $(|\Mset_c|-|\Mset_l|)$ is smaller than $2|\Mset_l|$. Therefore, in the next $|\Mset_c|-|\Mset_d|$ transmission slots, the base station transmits the remaining packets in $\Mset_c$ and the cooperating devices transmit $\frac{|\Mset_c|-|\Mset_d|}{2}$ packets from $\Mset_l$. At last, $|\Mset_l|-\frac{|\Mset_c|-|\Mset_d|}{2}$ packets are left from $|\Mset_l|$; it takes $\frac{2}{3}(|\Mset_l|-\frac{|\Mset_c|-|\Mset_d|}{2})$ transmission slots by the source and the cooperating devices, jointly to transmit these remaining packets. By summing the required number of transmission slots, the packet completion time for case (ii) is equal to $\frac{1}{3}(2|\Mset_l|+2|\Mset_c|+|\Mset_d|)$.

In addition, any packet in $\Mset_c$ contains one packet from the {\em Wants} set of each device by definition. On the other hand, any network coded packet in $\Mset_l$ contains one and only one packet from the {\em Wants} set of each device. Therefore, the inequality $|\Wset_n| \ge (|\Mset_c|+|\Mset_l|)$ holds for each device $n \in \Nset$ including the device with the minimum size of {\em Wants} set. Therefore, the following inequality, holds:

\begin{equation} \label{eq:M_l}
|\Mset_l| \leq (\min_{n \in \Nset} |\Wset_n|-|\Mset_c|).
\end{equation}

By using the above discussion, the maximum packet completion time under condition (2) is upper bounded by:
\begin{equation}
\begin{split}
T_{(2)}&=\frac{1}{3}(2|\Mset_l|+2|\Mset_c|+|\Mset_d|)\\
&\leq\frac{1}{3}(2\min_{n \in \Nset} |\Wset_n|+|\Mset_d|).
\end{split}
\end{equation}

\begin{example}
Let us consider three mobile devices with the {\em Wants} sets of $\Wset_A=\{p_1,p_2,p_5,p_8\}, \Wset_B=\{p_1,p_3,p_6,p_9,p_{11}\}, \Wset_C=\{p_1,p_4,p_7,p_{10},p_{11}\}$. By using Algorithm \ref{al:Groups}, $\Mset_c=\{p_1\}, \Mset_l=\{p_2+p_3+p_4,p_5+p_6+p_7,p_8+p_9+p_{10}\}, \Mset_d=\{p_{11}\}$. For this example, condition (2) is met; $(|\Mset_d|-|\Mset_l|)<|\Mset_c|<(|\Mset_d|+2|\Mset_l|)$. Accordingly, $3$ transmission slots are required; in the first transmission slot, $p_1$ is transmitted from the base station and at the same time $p_{11}$ is transmitted from device $A$. In the second transmission slot, $p_2+p_3+p_4$ is transmitted from the base station and $p_9+p_{10}$ is transmitted from device $A$. In the third transmission slot, $p_5+p_6+p_7$ is transmitted from the base station and $p_8$ is transmitted from device $B$.
\hfill $\Box$
\end{example}

\item{$|\Mset_c| \leq (|\Mset_d|-|\Mset_l|)$}\\
Under this condition, the base station starts transmitting the packets in $\Mset_c$; meanwhile in the local area the packets in $\Mset_d$ are transmitted from $x$ (the device that has all packets in $\Mset_d$). After $|\Mset_c|$ transmission slots, all packets in $\Mset_c$ have been transmitted by the base station and $|\Mset_d|-|\Mset_c|$ packets are left from $\Mset_d$. In the next $|\Mset_l|$ transmission slots, the base station transmits all packets in $\Mset_l$ and the cooperating devices transmit $|\Mset_l|$ packets from $\Mset_d$. At last, $|\Mset_d|-|\Mset_c|-|\Mset_l|$ packets are left from $\Mset_d$; it takes $\frac{|\Mset_d|-|\Mset_c|-|\Mset_l|}{2}$ transmission slots by the source and the cooperating devices, jointly to transmit these remaining packets. By summing the required number of transmission slots and from \ref{eq:M_l}, the maximum packet completion time under condition (3) is upper bounded by:

\begin{equation}
\begin{split}
T_{(3)}&=\frac{|\Mset_d|+|\Mset_c|+|\Mset_l|}{2}\\
&\leq\frac{|\Mset_d|+\min_{n \in \Nset} |\Wset_n|}{2}
\end{split}
\end{equation}

\begin{example}
Let us consider three mobile devices with the {\em Wants} sets of $\Wset_A=\{p_1,p_2\}$, $\Wset_B=\{p_1,p_3,p_5,p_6,p_9\}$, $\Wset_C=\{p_1,p_4,p_5,p_7,p_8,p_{10}\}$. By using Algorithm \ref{al:Groups}, $\Mset_c=\{p_1\}$, $\Mset_l=\{p_2+p_3+p_4\}$, $\Mset_d=\{p_5,p_6+p_7,p_8+p_9,p_{10}\}$. For this example, condition (3) is met; $|\Mset_c|<(|\Mset_d|-|\Mset_l|)$. Accordingly, $3$ transmission slots are required; in the first transmission slot, $p_1$ is transmitted from the base station and at the same time $p_{10}$ is transmitted from device $A$. In the second transmission slot, $p_2+p_3+p_4$ is transmitted from the base station and $p_8+p_9$ is transmitted from device $A$. In the third transmission slot, $p_5$ is transmitted from the base station and $p_6+p_7$ is transmitted from device $A$.
\hfill $\Box$
\end{example}

\end{enumerate}

By combining the packet completion time obtained from conditions (1), (2), and (3), the upper bound of $T_{upper,1}=\lceil \mathlarger{\max}(|\Mset_c|,\frac{1}{3}(2 \min_{n \in \Nset} |\Wset_n|+|\Mset_d|), \frac{1}{2}(\min_{n \in \Nset} |\Wset_n|+|\Mset_d|)) \rceil$ is achieved.

In addition, according to Algorithm \ref{al:Groups}, first vector $\boldsymbol v_m[n]$ is defined for each packet $p_m \in \Mset$. Then, different vectors may be combined as a network coded packet that can be transmitted from the source or the cooperating devices. In the worst case scenario, the vectors can not be combined as network coded packets. Therefore, we have at most $M=|\Mset|$ vectors, each representing one of the packets from set $\Mset$. In this case, each vector can be transmitted by a single transmission from the source or by a single transmission from the cooperating devices (if the packet represented by the vector is available in one of the devices). Under this worst case scenario, the upper bound of $T_{upper,2}=\lceil \mathlarger{\max}(|\Mset_c|,\frac{M}{2}) \rceil$ is achieved.

By combining $T_{upper,1}$ and $T_{upper,2}$, the packet completion time in Theorem \ref{theorem:better_upper} is obtained. This concludes the proof. 

\section*{\label{appendix_upper_batch_lossy}Appendix C: Proof of Theorem~\ref{th:upper_bound_batch_lossy}}
In \ncrn ~the packets are selected such that the delivered packets carry the most information about the missing packets in the local area. Therefore, the packets with the maximum average number of successful receivers are selected to be transmitted from the source and in the local area. Any other packet selection method with less average number of successful receivers delivers less information about the missing packets to devices and thus requires larger packet completion time. To find an upper bound on \ncrn, we consider one of such methods, where the packet selection at each transmission slot is as follows: (i) the source transmits a random linear combination of the missing packets in all devices and (ii) in the local area, the device with the maximum size of {\em Has} set is selected as the transmitter and transmits a linear combination of the packets in its {\em Has} set. If there are multiple of such devices, they are selected alternatively as the transmitter for the rest of the transmission slots until the end of the packet completion time. As seen, the packet selection in the source for this method is the same as \ncrn. But the packet selection in the local area differs from \ncrn; in \ncrn ~the packet with the maximum average number of successful receivers is selected to be transmitted, while in the mentioned method, there is no guarantee that the selected packet is associated with the maximum average number of successful receivers. Therefore, the packet completion time obtained from this mentioned method is an upper bound on the packet completion time for \ncrn. In the following, we consider the worst case scenario for the considered method and calculate the average packet completion time for this scenario. The calculated packet completion time is an upper bound on the packet completion time
 achieved by \ncrn.

At the beginning, device $x=\arg\max_{n \in \Nset} |\Hset_n|$ is selected as the transmitter in the local area. Therefore, any other device $n \in (\Nset \setminus x)$ receives two transmissions; one from the source with the probability of channel loss of $\eta_n$ and another from $x$ with the probability of channel loss of $\epsilon_{x,n}$. Device $x$ receives only one transmission which is transmitted from the source with the probability of channel loss of $\eta_x$. Therefore, in average, the size of $\Wset_x$ is reduced by $(1-\eta_x)$ and the size of $\Wset_n, n \in (\Nset \setminus x)$ is reduced by $(1-\eta_n+1-\epsilon_{x,n})=(2-\eta_n-\epsilon_{x,n})$ at each transmission slot until one of the devices other than $x$ is selected as the transmitter. According to the method, a device is selected as the transmitter if it has the largest {\em Has} set (or smallest {\em Wants} set) among all devices. Therefore, $x$ remains the transmitter until the size of {\em Wants} set for one of the other devices is equal to the size of {\em Wants} set for $x$. After the average of $k$ transmission slot, the size of $\Wset_n, n \in (\Nset \setminus x)$ is equal to $|\Wset_n|-k(2-\eta_n-\epsilon_{x,n})$ and the size of $\Wset_x$ is equal to $|\Wset_x|-k(1-\eta_x)$. Therefore, by considering device $r$ as $\arg\max_{n \in (\Nset \setminus x)} \frac{|\Wset_n|}{2-\eta_n-\epsilon_{x,n}}$, it takes at most $k = \frac{|\Wset_r|-|\Wset_x|}{(2-\eta_r-\epsilon_{x,r})-(1-\eta_x)}$ transmission slots ($k$ is obtained by solving the equation $|\Wset_x|-k(1-\eta_x)=|\Wset_r|-k(2-\eta_r-\epsilon_{x,r})$) for another device (which is device $r$) to be selected as the transmitter in average. On the other hand, it takes at most $\frac{|\Wset_x|}{1-\eta_x}$ transmission slots for device $x$ to receive all required packets and the size of its {\em Has} set becomes equal to $M$. We consider two cases :

\begin{enumerate}

\item $k \geq \frac{|\Wset_x|}{1-\eta_x}$

After at most $\frac{|\Wset_x|}{1-\eta_x}$ transmission slots, the size of {\em Has} set for device $x$ is equal to $M$ and the size of {\em Has} set for device $r$ is less than $M$. Thus $x$ remains the transmitter for the next transmission slots until the end of packet completion time. Therefore, the packet completion time is at most equal to the number of transmission slots required by $r$ to be satisfied; $T \leq \lceil \frac{|\Wset_r|}{2-\eta_r-\epsilon_{x,r}} \rceil$.

\item $k \leq \frac{|\Wset_x|}{1-\eta_x}$

After at most $k$ transmission slots, the size of {\em Has} set for device $r$ is equal to the size of {\em Has} set for device $x$. Therefore, according to the method, these devices are selected as the transmitter alternatively for the rest of the transmission slots until the end of the packet completion time. In this way, the sizes of $\Wset_r$ and $\Wset_x$ are reduced by $(1-\eta_r)+(2-\eta_r-\epsilon_{x,r})$ and $(1-\eta_x)+(2-\eta_x-\epsilon_{r,x})$ in every two consecutive transmission slots after the first $k$ transmission slots, respectively and thus it takes at most $T_r=k+2\frac{|\Wset_r|-k(2-\eta_r-\epsilon_{x,r})}{3-2\eta_r-\epsilon_{x,r}}$ and $T_x=k+2\frac{|\Wset_x|-k(1-\eta_x)}{3-2\eta_x-\epsilon_{r,x}}$ transmission slots for devices $r$ and $x$, respectively to be satisfied. By replacing $k=\frac{|\Wset_r|-|\Wset_x|}{(2-\eta_r-\epsilon_{x,r})-(1-\eta_x)}$ in the obtained expression for $T_x$ and $T_r$, equations \ref{eq:Tx} and \ref{eq:Tr} is obtained. The total required number of transmission slots is equal to maximum of $T_x$ and $T_r$.

\end{enumerate}

In addition, each packet in $\Mset_c$ can only be transmitted from the source until at least one of the devices can receive it successfully. Each packet that is transmitted from the source is received successfully by at least one of the devices with probability of $1-\prod_{n \in \Nset} \eta_n$. Therefore, it takes $\frac{1}{1-\prod_{n \in \Nset} \eta_n}$ transmissions for each packet in $\Mset_c$ to be received successfully by at least one of the devices and thus the number of transmission slots cannot be less than $\lceil \frac{|\Mset_c|}{1-\prod_{n \in \Nset} \eta_n} \rceil$. By considering this fact and the results from cases (1) and (2), the upper bound in Theorem~\ref{th:easy_upper_bound} is obtained. This concludes the proof.

\section*{\label{appendix_upper_instant_lossy}Appendix D: Proof of Theorem~\ref{th:upper_bound_lossy}}
We consider the initial sets of $\Mset_c$, $\Mset_l$, and $\Mset_d$ constructed for the first transmission slot of \ncin. The calculated number of transmission slots to transmit all the packets in these sets gives an upper bound on the packet completion time by \ncin. The reason is that in \ncin, the packet completion time is improved by updating the sets $\Mset_c$, $\Mset_l$, and $\Mset_d$ at each transmission slot and thus the resulted packet completion time from \ncin ~is less than the packet completion time resulted from \ncin ~without updating the sets $\Mset_c$, $\Mset_l$, and $\Mset_d$ at each transmission slot (\ie these sets are set to their initializations at the beginning of the first transmission slot).
In the following, we first derive the expressions for the average packet completion times for the source to transmit the packets in $\Mset_c$, $\Mset_l$, and $\Mset_d$, denoted by $T_{s,c}$ (Eq. \ref{eq:T_sc}), $T_{s,l}$ (Eq. \ref{eq:T_sl}), and $T_{s,d}$ (Eq. \ref{eq:T_sd}), respectively and the expressions for the average packet completion times for the devices in the local area to transmit the packets in $\Mset_l$, and $\Mset_d$, denoted by $T_{l,l}$ (Eq. \ref{eq:T_ll}), and $T_{l,d}$ (Eq. \ref{eq:T_ld}), respectively. Then, we calculate the average packet completion time when \ncin, without updating the sets $\Mset_c$, $\Mset_l$, and $\Mset_d$ at each transmission slot, is used to recover the missing packets in all devices.

The average number of transmissions from the transmitter (or the source) required to satisfy a device is equal to $1/(1-\epsilon)$, where $\epsilon$ is the link loss between the transmitter (or the source) and the device. The average number of transmissions required to satisfy a set of targeted receivers is restricted by the required number of transmissions for the receiver with the maximum channel loss. This proves Eqs. \ref{eq:T_sc}, \ref{eq:T_sl}, \ref{eq:T_sd}, and \ref{eq:T_ld}. To transmit packet $p$ in $\Mset_l$ from the cooperating devices in the local area, first device $t$ (with the maximum average number of successful receivers) is selected among all devices to transmit a partial of the packet. The set of targeted receivers for this transmission is $\{n \mid (v_p[n] \neq v_p[x])\}$. Therefore, it takes an average of $\frac{1}{1-max_{n \mid (v_p[n] \neq v_p[x])} \epsilon_{x,n}}$ transmissions to transmit partial of packet $p$ (the first term in Eq. \ref{eq:T_ll}). Then, the residual part of $p$, which includes the uncoded packet in $p$ that is wanted by device $x$ ($v_p[x]$), is transmitted. This packet is transmitted from, $x'$, the device with the maximum number of successful receivers, and targets the devices in the set $\{n \mid (v_p[n] = v_p[x])\}$). Therefore, it takes an average of $\frac{1}{1-max_{n \mid (v_p[n] = v_p[x])} \epsilon_{x',n}}$ transmissions to transmit residual of packet $p$ (the second term in Eq. \ref{eq:T_ll}). This proves Eq. \ref{eq:T_ll}.

To prove Eq. \ref{eq:b2_random_l}, we first consider two cases based on the relative packet completion time for transmitting the packets in the sets $\Mset_c$, $\Mset_l$, and $\Mset_d$ and then calculate the average packet completion time obtained from each of these cases.

\begin{enumerate}
\item{$T_{s,c} \geq (T_{l,l}+T_{l,d})$}\\
Under this condition, the source starts transmitting the packets in $\Mset_c$, which takes the average of $T_{s,c}$ transmissions. Meanwhile the packets in the set $\Mset_n=\Mset_l \cup \Mset_d$ are transmitted in the local area, which takes the average of $T_{l,l}+T_{l,d}$ transmissions. After $T_{l,l}+T_{l,d}$ transmission slots, all the packets in $\Mset_d$ and $\Mset_l$ are transmitted in the local area and the average of $|\Mset_c|-\frac{|\Mset_c|(T_{l,l}+T_{l,d})}{T_{s,c}}=\frac{|\Mset_c|(T_{s,c}-T_{l,l}-T_{l,d})}{T_{s,c}}$ packets are left from $\Mset_c$; it takes an average of $\frac{|\Mset_c|(T_{s,c}-T_{l,l}-T_{l,d})}{T_{s,c}} \times \frac{T_{s,c}}{|\Mset_c|}=T_{s,c}-T_{l,l}-T_{l,d}$ transmissions for the source to transmit these packets. By summing the required number of transmission slots, the average packet completion time under condition (1) is equal to:

\begin{equation}
T_{(1)}=T_{s,c}
\end{equation}

\item{$T_{s,c} \leq (T_{l,l}+T_{l,d})$}\\
Under this condition, the base station starts transmitting the packets in $\Mset_c$; meanwhile the packets in the set $\Mset_n=\Mset_l \cup \Mset_d$ are transmitted in the local area. Since $T_{s,c} \leq (T_{l,l}+T_{l,d})$, after $T_{s,c}$ transmission slots, all the packets in $\Mset_c$ have been transmitted by the source and the average of $|\Mset_d|+|\Mset_l|-\frac{T_{s,c}(|\Mset_l|+|\Mset_d|)}{T_{l,l}+T_{l,d}}=\frac{(T_{l,l}+T_{l,d}-T_{s,c})(|\Mset_l|+|\Mset_d|)}{T_{l,l}+T_{l,d}}$ packets are left from $\Mset_n$; It takes the average of $\frac{(T_{l,l}+T_{l,d}-T_{s,c})(|\Mset_l|+|\Mset_d|)}{T_{l,l}+T_{l,d}} \times \frac{(T_{l,l}+T_{l,d})(T_{s,l}+T_{s,d})}{(T_{l,l}+T_{l,d})+(T_{s,l}+T_{s,d})} \times \frac{1}{|\Mset_l|+|\Mset_d|}=\frac{(T_{s,l}+T_{s,d})(T_{l,l}+T_{l,d}-T_{s,c})}{(T_{l,l}+T_{l,d}+T_{s,l}+T_{s,d})}$ transmission slots to transmit these packets by using both the source and the cooperating devices in the local area. By summing the required number of transmission slots, the packet completion time under condition (2) is equal to:

\begin{align*}
  & T_{(2)}= T_{s,c}+\frac{(T_{s,l}+T_{s,d})(T_{l,l}+T_{l,d}-T_{s,c})}{T_{l,l}+T_{l,d}+T_{s,l}+T_{s,d}} \\
  = & \frac{T_{s,c}(T_{s,l}+T_{s,d})+T_{s,c}(T_{l,l}+T_{l,d})}{T_{l,l}+T_{l,d}+T_{s,l}+T_{s,d}}+\\
  &\frac{(T_{s,l}+T_{s,d})(T_{l,l}+T_{l,d})-T_{s,c}(T_{s,l}+T_{s,d})}{T_{l,l}+T_{l,d}+T_{s,l}+T_{s,d}}\\
  = & \frac{T_{s,c}(T_{l,l}+T_{l,d})+(T_{s,l}+T_{s,d})(T_{l,l}+T_{l,d})}{T_{l,l}+T_{l,d}+T_{s,l}+T_{s,d}}\\
  = & \frac{(T_{l,l}+T_{l,d})(T_{s,c}+T_{s,l}+T_{s,d})}{T_{l,l}+T_{l,d}+T_{s,l}+T_{s,d}}.
\end{align*}

\end{enumerate}

By combining the packet completion time obtained from conditions (1) and (2), the upper bound presented in Theorem \ref{th:upper_bound_lossy} is achieved. This concludes the proof.

\section*{\label{appendix_lower}Appendix E: Proof of Theorem~\ref{th:lowerbound_loss}}
To prove Eq. \ref{eq:lower_loss}, we first derive a lower bound on the packet completion time to recover the missing packets in device $n$, when the cellular and D2D links are used, jointly. This term is denoted by $T_n$. In the best case scenario, device $n$ receives two simultaneous transmissions at each transmission slot, one from the source with the loss probability of $\eta_n$ and another one from the transmitter device $x$ with the loss probability of $\epsilon_{x,n}$. Therefore, the average number of packets that device $n$ receives at each transmission slot, is equal to $(1-\eta_n)+(1-\epsilon_{x,n})=2-\eta_n-\epsilon_{x,n}$ and thus, in the best case scenario, it takes an average of $\frac{|\Wset_n|}{2-\eta_n-\epsilon_{x,n}}$ transmission slots to satisfy device $n$ with the transmitter device $x$ for the transmission in the local area. Again, in the best case scenario, $x$ is selected such that $T_n$ is minimized. Therefore, we have:

\begin{equation}
\begin{split}
T_n &\geq \frac{|\Wset_n|}{2-\eta_n-\epsilon_{x,n}}\\
&\geq \min_{x \in (\Nset \setminus n)} \frac{|\Wset_n|}{2-\eta_n-\epsilon_{x,n}}.
\end{split}
\end{equation}

By using network coding algorithms, the packet completion time to satisfy all devices is equal to maximum packet completion time required to satisfy each device. In other words, we have:

\begin{equation}
\begin{split}
T &= \max_{n \in \Nset} T_n\\
&\geq \max_{n \in \Nset} (\min_{x \in (\Nset \setminus n)}\frac{|\Wset_n|}{2-\eta_n-\epsilon_{x,n}}).
\end{split}
\end{equation}

On the other hand, the packets in $\Mset_c$ can only be sent through the cellular link, because they are not available in any of the devices and thus cannot be transmitted through D2D links. The average of required number of transmission slots for a packet in $\Mset_c$ to be received successfully by at least one of the devices (so that it will be available to be transmitted through D2D links) is equal to $\frac{1}{1-\prod_{n \in \Nset} \eta_n}$. Therefore, the minimum packet completion time should be larger than $\frac{|\Mset_c|}{1-\prod_{n \in \Nset} \eta_n}$. Thus, the packet completion time is bounded by $T \geq \max (\frac{|\Mset_c|}{1-\prod_{n \in \Nset}\eta_n},\max_{n \in \Nset}(\min_{x \in (\Nset \setminus n)}\frac{|\Wset_n|}{2-\eta_n-\epsilon_{x,n}})$. Furthermore, since the packet completion time can only have an integer value, the packet completion time is lower bounded by $\lceil \max (\frac{|\Mset_c|}{1-\prod_{n \in \Nset}\eta_n},\max_{n \in \Nset}(\min_{x \in (\Nset \setminus n)}\frac{|\Wset_n|}{2-\eta_n-\epsilon_{x,n}}) \rceil$. This concludes the proof.

\section*{\label{appendix_lower}Appendix F: Signalling Overhead Analysis}
In our proposed algorithms, at each transmission slot, two network coded packets are selected to be transmitted from the two interfaces of the cellular and D2D. The decision of which network coded packet to be transmitted from the cellular links is made by the source. Meanwhile, the decisions of which network coded packet and which device to be selected as the transmitter in the local area, are made by the controller (which is selected randomly among mobile devices). In order for the source and the controller to make these decisions, the information about the packets that each device has received successfully, needs to be transferred to the source and the controller. Therefore, the signaling overhead at each transmission slot is the sum of (i) $O_d$ bits for sending the controller's decision to the selected transmitter device (the controller needs to inform the transmitter device about its decision of packet transmission through D2D), (ii) $O_{nc}$ bits as network coding overhead (the coefficients of uncoded packets in a network coded packet should be included as header in the transmitted network coded packet), and (iii) $O_{ack}$ bits for sending acknowledgment packets from the targeted receivers at the end of the transmission to the source and controller indicating successful/unsuccessful transmissions. The signaling overhead $O_d$ is related to the packet transmission through D2D links and the signaling overheads related to $O_{nc}$ and $O_{ack}$ are related to the packet transmissions through both cellular and D2D links. We analyze the signaling overhead for our proposed methods, lossy and lossless \ncrn ~and \ncin, in detail in the following.

{\bf Stage One:} In stage one, we use the single interface of cellular links for packet transmissions. Therefore, there is no packet transmissions in the local area, so $O_d=0$. $O_{nc}$ is also equal to $0$ as the uncoded packets are transmitted to all devices without network coding in stage one. At the end of each packet transmission, acknowledgment packets need to be transmitted from each device to both source and controller to provide them with the required information for making packet transmission decisions in stage two. Therefore, $O_{ack}$ for each packet transmission and each user is equal to $2$ bits ($1$ bit for sending the acknowledgment packet to the source and one more bit for sending the acknowledgment packet to the controller). Thus, the overhead fraction per packet transmission to each user is equal to $\frac{O_d + O_{nc} + O_{ack}}{P} = \frac{2}{P}$, where $P$ is the size of each uncoded packet in bits.

{\bf Stage Two, Lossless NCMI-Batch:} For lossless \ncrn, at each transmission slot, the controller selects one of the devices as the transmitter to send a random linear combination of the packets in its {\em Has} set; a packet with the size of one bit is sufficient to be transmitted from the controller to the selected transmitter to inform the transmitter to send a packet; $O_d=1$ bit. Then, two random linear network coded packets need to be transmitted; one from the source and another from the selected transmitter. For each random linear network coded packet, we need to include the coefficients of each uncoded packet as the packet header; e.g. a random linear combination of the packets $p_1,p_2,...,p_M$ is equal to the network coded packet $p=a_1p_1+a_2p_2+...+a_Mp_M$. For transmitting the network coded packet $p$, all the coefficients of $a_i, i=1,2,...,M$ are added to the data packet $p$ as a packet header. The number of bits required for displaying each coefficient is equal to $log(F)$, where $F$ is the size of the field from which the coefficients are selected. Therefore, $O_{nc}$ is equal to $Mlog(F)$ bits for sending each network coded packet and $2Mlog(F)$ bits for sending the two network coded packets via the two interfaces of cellular and D2D. Finally, as the channels are lossless, all packets will be received successfully at the targeted receivers and there is no need to send acknowledgment packets at the end of each packet transmissions; $O_{ack}=0$. Therefore, the overhead fraction per transmission slot is equal to $\frac{O_d + O_{nc} + O_{ack}}{2P}  = \frac{1+2Mlog(F)}{2P}$ bits for lossless \ncrn. Note that the multiplier $2$ in the denominator is due to sending two network coded packets at each transmission slots over the two interfaces, each with the size of $P$.

{\bf Stage Two, Lossless NCMI-Instant:} For lossless \ncin, at each transmission slot, the controller selects one of the devices as the transmitter and determines the set of packets to be XORed as the instantly decodable network coded packet and transmitted from the transmitter. In order for the controller to inform the transmitter which packets should be XORed, a packet with the size of $M$ bits ($M$ is the number of the missing packets in all devices) is required; if the $i$th bit is $1$, $p_i$ should be included in the IDNC packet and if it is equal to $0$, $p_i$ should not included in the IDNC packet. For example, assume that the controller determines the IDNC packet $p=p_1+p_3+p_4$ for the set of missing packets $\{p_1,p_2,p_3,p_4,p_5\}$, with $M=5$ packets, to be transmitted from the selected transmitter. Then, the controller needs to send the packet containing $O_d=5$ bits of $[1 0 1 1 0]$ to the transmitter to inform it about its decision. Therefore, $O_d=M$ bits for \ncin. Then, the transmitter and the source need to send IDNC packets, in which the coefficients of each uncoded packet should be included as the header. As an IDNC packet is created by XORing the uncoded packets, $1$ bit is sufficient to represent the coefficient of each uncoded packet, so $M$ bits are required for the coefficients of all uncoded packets. Therefore, the total overhead due to sending network coding coefficients of the two IDNC packets, transmitted over cellular and D2D, is equal to $O_{nc}=2M$ bits, at each transmission slot. Finally, since the channels are lossless, $O_{ack}=0$, as discussed in the previous paragraph. Therefore, the overhead fraction per transmission slot is equal to $\frac{O_d + O_{nc} + O_{ack}}{2P}  = \frac{M+2M}{2P}=\frac{3M}{2P}$ bits for lossless \ncin.

{\bf Stage Two, Lossy NCMI-Batch:} The analysis of $O_d$ and $O_{nc}$ for lossy \ncrn, is the same as lossless \ncrn; $O_d=1$ bit and $O_{nc}=2Mlog(F)$ bits. However, since the channels are lossy here, at the end of each packet transmission, acknowledgment packets need to be transmitted from the targeted receivers to both source and controller to provide them with the information for making future packet transmission decisions. Therefore, for each packet transmission, the size of acknowledgment packet is equal to $2$ bits ($1$ bit for sending the acknowledgment packet to the source and one more bit for sending the acknowledgment packet to the controller) for each targeted receiver. Since, at each transmission slot, two packets are transmitted, the size of overhead due to sending acknowledgement packets is $4$ bits per each targeted receiver. Considering $N_t$ as the number of targeted receivers, the total size of overhead due to sending acknowledgment packets, is equal to $O_{ack}=4N_t$ bits, at each transmission slot. Therefore, the overhead fraction at each transmission slot is equal to $\frac{O_d + O_{nc} + O_{ack}}{2P}  = \frac{1+2Mlog(F)+4N_t}{2P} \leq \frac{1+2Mlog(F)+4N}{2P}$ for lossy \ncrn. Note that here we assume that the acknowledgement packets are not lost, for the sake of simplicity. If an acknowledgement packet is lost, it will be retransmitted and thus the overhead will be slightly increased.

{\bf Stage Two, Lossy NCMI-Instant:} The analysis of $O_d$ and $O_{nc}$ for lossy {\tt NCMI-Instant} is the same as lossless \ncin; $O_d=M$ bits and $O_{nc}=2M$ bits. Also, the analysis $O_{ack}$ for lossy \ncin ~is the same as lossy \ncrn; $O_{ack}=4N_t$ bits. Therefore, the overhead fraction at each transmission slot is equal to $\frac{O_d + O_{nc} + O_{ack}}{2P}  = \frac{M+2M+4N_t}{2P}=\frac{3M+4N_t}{2P} \leq \frac{3M+4N}{2P}$ for lossy \ncin.

In the following, we give an example on the calculation of overhead fraction for stage one and stage two of \ncrn ~and \ncin.

\begin{example}
Assume that the packet size of each transmitted packet is $P=1000$ Bytes $=8000$ bits, the field size is $F=256$ for \ncrn, the number of missing packets is $M=30$ and the number of devices is $N=5$. The overhead fraction for each packet transmission in stage one is equal to $\frac{2}{P}=0.00025$. The overhead fraction at each transmission slot for lossless \ncrn, lossless \ncin, lossy \ncrn, and lossy \ncin ~at stage two is equal to $\frac{1+2Mlog(F)}{2P}=0.03006$, $\frac{3M}{2P}=0.00562$, $\frac{1+2Mlog(F)+4N_t}{2P} \leq 0.03131$, and $\frac{3M+4N_t}{2P} \leq 0.00687$, respectively.''
\hfill $\Box$
\end{example} 

\end{document}